\documentclass[fleqn,usenatbib]{mnras}

\newcommand{\kms}{{\rm {\,km\, s^{-1}}}}

\newcommand{\gauss}{{\rm G}}

\usepackage{academicons}
\usepackage{orcidlink}
\usepackage{newtxtext,newtxmath}

\usepackage[T1]{fontenc}
\usepackage{float}
\usepackage{pdflscape}

\DeclareRobustCommand{\VAN}[3]{#2}
\let\VANthebibliography\thebibliography
\def\thebibliography{\DeclareRobustCommand{\VAN}[3]{##3}\VANthebibliography}

\usepackage{graphicx}	%
\usepackage{amsmath}	%
\usepackage{tikz,xcolor,hyperref}
\usepackage{tabularx}
\usepackage{siunitx}

\usepackage{case-symbols}
\usepackage{orcidlink} %

\title[High-latitude CMEs in AB Dor]{High-latitude coronal mass ejections on the young solar-like star AB Dor}

\author[Strickert, Evensberget \& Vidotto]{
    K. M. Strickert \orcidlink{0000-0001-6584-5969}, %
    D. Evensberget \orcidlink{0000-0001-7810-8028}\thanks{E-mail: evensberget@strw.leidenuniv.nl},
    A.~A. Vidotto \orcidlink{0000-0001-5371-2675}
\\
Leiden Observatory, Leiden University, PO Box 9513, 2300 RA Leiden, The Netherlands
}

\date{Accepted XXX. Received YYY; in original form ZZZ}

\pubyear{2024}

\begin{document}
\label{firstpage}
\pagerange{\pageref{firstpage}--\pageref{lastpage}}
\maketitle

\begin{abstract}
AB
Dor is a young solar-type star with a surface large-scale magnetic field
$10^2$ to $10^3$ times
stronger than the that of the Sun.
Although strong magnetic fields are thought to inhibit coronal mass ejections (CMEs), dimming signatures typically associated with an eruptive CME were recently observed in AB Dor. The uninterrupted, long-duration dimming signal suggests that a CME took place at a high latitude, where it remained in view as the star rotates.
A high-latitude CME is also consistent with observations that indicate that AB Dor hosts polar active regions.
To investigate magnetic confinement in AB Dor, we conduct a parametric modelling study of twenty-one CMEs at latitudes $\sim 60^\circ$, varying the location, mass and magnetic field strength of an injected flux rope. Twelve models had the flux rope located in an open magnetic field region, while the remaining nine were in a closed region.
Results show that CMEs in open-field regions are in general more likely to erupt.
The four eruptive CMEs from closed regions had high free magnetic energies $\gtrsim 3\times 10^{35}$~erg, and ten CMEs, predominantly from the closed-field regions (8/10) were confined. CMEs in closed-field regions exhibited lower kinetic energies, since part of the CME energy was expended to overcome magnetic tension and break open the overlying field.
In conclusion our work suggests that eruptive CMEs in AB Dor may occur in high-latitude regions of open magnetic field, as the magnetic tension in such regions does not significantly inhibit the eruption.

\end{abstract}


\section{Introduction}

Even though the Sun is a relatively magnetically inactive  star, it hosts a plethora of burst and transient events, in which free magnetic energy is violently released and converted into radiation and the acceleration of particles. The sudden burst of radiation in the form of flares can be associated to the expelling of material from the Sun in the form of filament/prominence eruptions and coronal mass ejections (CMEs). The flare-CME association is in particular more evident for large solar flares \citep[e.g.,][]{2006ApJ...650L.143Y}.
Given the magnetic nature of these explosive events, and the fact that solar-like stars can be substantially more magnetically active than the Sun (particularly    at young ages, e.g., \citealt{1999MNRAS.302..437D, 2015MNRAS.449....8W, 2020A&A...643A..39C}), solar-like stars are expected to not only host bursty events, but even more frequent and energetic ones \citep{2012ApJ...760....9A, 2013ApJ...764..170D, 2015ApJ...809...79O, 2019MNRAS.482.2853J, 2020MNRAS.494.3766O}. This idea is supported by the detection of stellar flares and super-flares \citep[e.g.][]{2004A&A...416..713G, 2012Natur.485..478M, 2014ApJ...797..121H, 2024LRSP...21....1K}.

In spite of the numerous detections of stellar flares observed across the electromagnetic spectrum, unambiguous detection of a counterpart CME remains challenging \citep{2019ApJ...877..105M, 2022arXiv221105506N}.
The indirect evidence of the presence of stellar CMEs is based on a range of detection methods, including Doppler shifts in Balmer lines \citep{2019A&A...623A..49V}, X-ray blueshifted emission \citep{2019NatAs...3..742A,2022ApJ...933...92C} and, more recently, coronal dimming observed in the X-ray, extreme ultra-violet (EUV) and far ultra-violet \citep{2020MNRAS.493.4570L, 2021NatAs...5..697V}.\footnote{Although Type II radio bursts, usually associated to solar CMEs, are promising ways to detect stellar CMEs \citep{2016ApJ...830...24C}, no stellar type II radio bursts have been reported yet.}
Eruptions have also recently been reported in the young, rapidly rotating
    star EK~Draconis~\citep{2022NatAs...6..241N,2024ApJ...961...23N}
    and in \(\epsilon\)~Eridani~\citep{2022ApJ...936..170L}.
In the Sun, coronal dimming is usually observed after eruptive events \citep{2016SoPh..291.1761H, 2016ApJ...830...20M}, with more than 80\% of solar CME events showing succeeding dimming signatures \citep{2021NatAs...5..697V}. As the CME propagates, evacuated mass behind the CME shock front do not contribute (or contribute less) to X-ray and EUV coronal emission \citep{2016ApJ...830...20M}, generating a dimming in X-ray/EUV light curves.

Motivated by the solar analogy, it has been suggested that stellar CMEs could also produce dimming signatures. Recently, \citet{2021NatAs...5..697V} examined stellar spectra in EUV and X-ray from  observational archives for about 200 stars. These authors detected 13 different stars that showed dimming associated with flares---this association is indicative of stellar CMEs. In particular, these authors found five dimming events on AB Dor (HD36705), a rapidly rotating young solar-mass star, with an age of about 120~Myr, a rotation period of 0.51 days and kG surface magnetic fields \citep{2013ApJ...766....6B, 2009A&ARv..17..251S, 1999MNRAS.302..437D, 2007MNRAS.377.1488H}.

The coronal dimming event from 1994 recorded by EUVE showed a {surprisingly} long dimming lasting 13.3~h from flare peak to the end of the dimming \citet{2021NatAs...5..697V}. Because the duration of the event is comparable to the $\sim 0.5$~day rotation period of AB Dor,  \citet{2021NatAs...5..697V} suggested that the associated CME might have originated in a region that is constantly in view of the observer, namely at high latitudes. This is in line with observations that show that AB Dor's rotation axis has an inclination of 60$^\circ$ \citep{1999MNRAS.302..437D}. Interestingly, AB Dor exhibits polar active regions \citep{1994MNRAS.269..814C}.
Using again the Sun as analogy, given that the majority of solar CMEs tend to originate in active regions, it is very plausible that the CME event reported by \citet{2021NatAs...5..697V} could have also originate from polar active regions. In the Sun,
both low- and high-latitude CMEs are observed and their source regions differ. While the more abundant low-latitude CMEs originate from solar active regions (concentrated in a belt of $\pm 40^\circ$ around the equator), high-latitude CMEs  originate from polar crown filaments
\citep[e.g.,][]{2015ApJ...809..106G, 2022ApJ...932...62L}. Understandably, the spacial information we have available for AB Dor is substantially inferior than that of the Sun. Here, we speculate that the source of the high-latitude CMEs in AB Dor is its polar active regions.

One potential problem for the eruption of CMEs in AB Dor, or any other magnetically active star, is that the star itself hosts a strong ($\sim$~kG), large-scale surface magnetic field, that can prevent the eruption of CMEs. The numerical study of \citet{2018ApJ...862...93A} showed that a stronger overlying large-scale dipolar magnetic field can prevent CMEs from escaping, unless their  energies are much stronger than energies typically associated to solar CMEs.
To better understand the eruption (or confinement) of CMEs in the young, active star AB Dor, in this work we conduct  a parametric study of CME eruptions using three-dimensional magnetohydrodynamical models. The combination of a dimming time  that is longer than the rotation period of AB Dor \citep{2021NatAs...5..697V}  and the presence of  polar active regions \citep{1994MNRAS.269..814C}, led us to explore the ejection of high-latitude CMEs only.
Our paper is structured as follows: Section \ref{sec:models} introduces the numerical model we use in this work, including the data-driven background wind of AB Dor and the eruption model of the CMEs.
In Section \ref{sec.CME}, we present the physical properties of the our simulated CMEs of our numerical study, their morphology and their eruptive (or confined) nature, and their energetics. Section \ref{sec:discussion_and_conclusion} shows our discussion and conclusions of this work.

\section{Models}\label{sec:models}
\subsection{The background stellar wind model }\label{sec:AWSom_Model}

To model the wind of AB Dor, we use the Alfvén Wave Solar Model \textsc{awsom}, \citep{2013ApJ...764...23S,2014ApJ...782...81V} which is implemented in the 3D MHD code \textsc{BATS-R-US} \citep{1999JCoPh.154..284P, 2012JCoPh.231..870T}. This  model solves the magnetohydrodynamics (MHD) equations, assuming that Alfvén waves heat and accelerate the stellar wind. Although initially developed to describe the solar wind, this model has been used also to model winds of cool dwarf stars \citep[e.g.][]{2018ApJ...862...93A, 2021MNRAS.500.3438O, 2021MNRAS.504.1511K, Evensberget_Carter_Marsden_Brookshaw_Folsom_2021, evensberget2022, 2023MNRAS.524.2042E, 2023A&A...678A.152V}. We refer the reader to these papers for more details of the model.

\begin{figure}
    \centering
    \includegraphics[width = .46\textwidth]{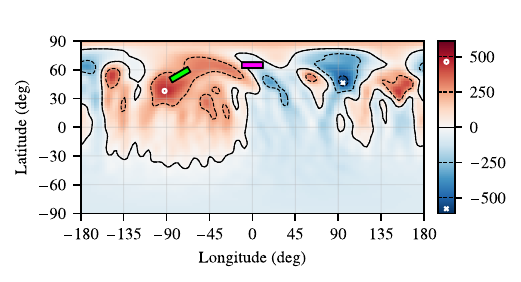}
    \caption{Surface magnetic field structure of AB Dor, derived from spectropolarimetric observations performed in December 2002 \citep{2007MNRAS.377.1488H}. Two rectangular patches display the two different flux rope locations for the simulated CME models: the green patch is located at open magnetic field region, while the magenta patch at closed-field region. The white circle and cross indicate the maximum and minimum magnetic field strengths. The average strength of the depicted radial field is \qty{105}{\gauss}.}
    \label{fig:magnetogram}
\end{figure}

In addition to the mass ($0.86~M_\odot$), radius ($0.96~R_\odot$), and rotation period ($0.514$~d) of AB Dor, the main input of our model is the radial surface magnetic field of AB Dor. Here, we use the December 2002 surface magnetic field map,  obtained from a series of spectropolarimetric observations, from \citet{2007MNRAS.377.1488H}. This map is reconstructed using Zeeman-Doppler Imaging~\cite[ZDI,][]{1989A&A...225..456S, 2009ARA&A..47..333D}. The map we adopt in our work use is described by spherical harmonics up to degree $\ell_\text{max} = 25$ and is shown in Figure \ref{fig:magnetogram}. We see that the radial surface field strengths range from $-573$ to $466\,$G.

The inner boundary of our simulation is placed in the chromosphere, where we adopt similar values as those used in the simulations of the solar wind by e.g.~\citet{2023MNRAS.524.2042E}. We assume a density of $2 \times 10^{-11}\,$cm$^{-3}$ and temperature of  $5 \times 10^4\,$K. The wave dissipation length scale at the boundary is assumed to be $1.5 \times 10^5$\,m\,$\sqrt{\text{T}}$ %
and the  Poynting flux-to-field ratio at the inner boundary is set to \qty{e6}{\watt\per\square\meter\per\tesla}. Finally, the heat conduction contains a stochastic heating term  of $0.18$, a heat flux parameter of $1.05$ and a collisionless radius of $ 5R_\star$. We refer the reader to \citet{2014ApJ...782...81V} for a detailed description of these model parameters.

Our numerical grid is spherical with the smallest grid size of $1/256 R_\star$ close to the stellar surface (radial distances $\leq 1.7\,R_\star$) and around the current sheet. We also have a conical region with the same resolution extending throughout the domain, having its base at the region where the CME is initiated and propagates.
Elsewhere in the grid, the resolution is kept at $1/64\,R_\star$. The grid extends out to $24\,R_\star$.

\begin{figure}
    \centering
    \includegraphics[width = \columnwidth]{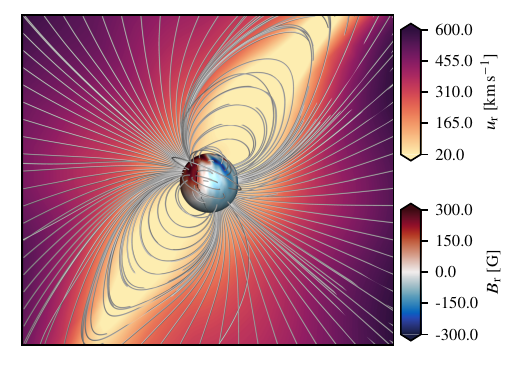}
    \caption{Steady-state wind solution of AB Dor, where we used the ZDI map based on the work of \citet{2007MNRAS.377.1488H} and presented in Figure \ref{fig:magnetogram}. The background is colour-coded by the radial wind velocity and the stellar surface by the radial magnetic field ZDI map. The domain of the figure extend out to roughly eight stellar radii and the axis of rotation is in the vertical direction. }
    \label{fig:Steady-steate_solution}
\end{figure}

Prior to the injection of the CME flux rope, we let the simulation evolve until the stellar wind reaches a steady-state configuration, as shown in Figure \ref{fig:Steady-steate_solution}.
At this point, we compute the wind mass-loss rate as the mass flux through a spherical surface at a certain distance above the stellar surface $\dot{M} = \oint \rho u_r\,\mathrm{d}A$.
We found a mass-loss rate of
$10^{-12} M_\odot/\textrm{yr} \sim 53\, \dot{\text{M}}_\odot$, which lies
within estimates of \citet{2010ApJ...721...80C} and below the upper limit derived from prominence studies by \citet{2019MNRAS.482.2853J}.
We see that the magnetic field embedded in the stellar wind resembles an inclined dipole, and we find wind speeds reaching a maximum value of $\sim 1000\,$km\,s$^{-1}$ at large distances in the polar regions.

The steady-state wind solution of AB Dor  serves as the background wind for the 21  CME events we simulate.

\subsection{The eruption model of the CME}\label{sec:flux_rope}

\begin{figure}
    \centering
    \includegraphics[width = \columnwidth]{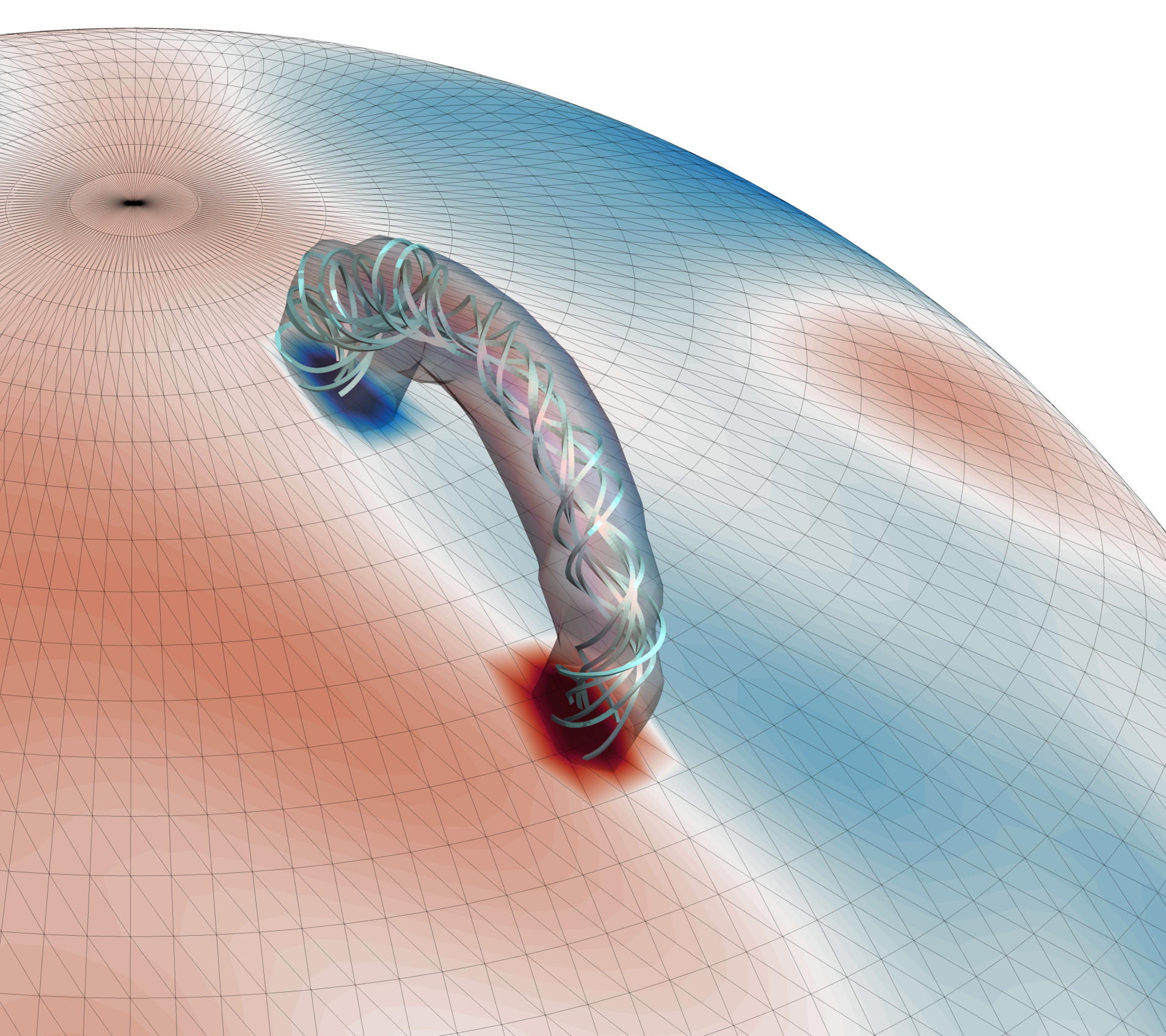}
    \caption{An example initial configuration of the flux rope created using the model of \citet{1999A&A...351..707T} with a flux rope mass of $3.5 \times 10^{18}\,$g and a central magnetic field strength of $1000\,$G. The colours at the inner boundary indicate the polarity of the radial magnetic field with values $-900\,$G (blue) to $+900\,$G (red). The underlying line current present during the formation of the flux rope (c.f.,\,\ref{sec:flux_rope}) generates the twisted magnetic field lines seen inside the flux rope in the figure.}
    \label{fig:Flux_rope}
\end{figure}

To initiate the CMEs in our models, we add a magnetic flux rope whose footpoints are anchored at the inner boundary of our simulations.
Similar CME eruption models have been adopted in numerical studies of solar \citep[e.g.][]{2013ApJ...773...50J, 2017ApJ...834..172J} and stellar \citep{2018ApJ...862...93A, 2022ApJ...924..115O} CMEs. Here, we use the flux rope model based on \citet{1999A&A...351..707T}, in which virtual magnetic charges and currents along the toroidal symmetry axis produce a toroidal flux rope with a spiralling magnetic field in its interior (see Figure \ref{fig:Flux_rope}).
The injection of the flux rope perturbs the background wind solution and in many cases this leads to an eruptive CME event.

To elucidate which flux rope parameters should be adopted in the simulations of stellar CMEs, our parametric study comprises 21 CME simulations with different flux rope characteristics.  The flux rope model by \citet{1999A&A...351..707T}  can be described by eight parameters.
The first three parameters describe the geometrical shape of the flux rope and are held constant in all our models: the major radius of the flux rope toroid $R_\text{maj} = 128\,$Mm, the minor radius $R_\text{min} = 20\,$Mm, and the depth below the inner boundary of the toroid symmetry axis $D = 1\,$Mm.
This loop structure mimics a strong magnetic bipole anchored at the surface of the star. In the context of solar CME models, the flux rope position and size are adjusted to match the solar active regions from which CMEs are observed to originate \citep[e.g.][]{2008ApJ...684.1448M, 2017ApJ...834..173J}. In the case of stars, these geometric considerations are not well constrained \citep{2023BAAS...55c.254L}; we cannot observationally resolve the size and distances of magnetic bipolar regions in stellar surfaces. Therefore, the values used here are the same as those used in  CME simulations of $\epsilon$ Eridani performed by \citet{2022ApJ...924..115O} and on the same ballpark as to values used in solar \citep[e.g.][]{2003ApJ...588L..45R, 2011JGRA..116.4106L, 2013ApJ...773...50J} and stellar \citep[e.g.,][]{2020ApJ...895...47A} CME simulations.

The next two parameters determine the energy of the resulting CME and are varied in our simulations: the flux rope mass and magnetic field. In our work, $M_\textsc{fr}$ ranges from \qtyrange{3.5e14}{3.5e18}{\gram}. The lower end of these values is comparable to the flux rope mass of~\citet{2020ApJ...895...47A}.
Given these assumed masses and the volume of the flux rope (\qty{\sim 5e+29}{\cubic\centi\meter}, according to the assumed geometry), the average density inside the flux rope ranges from $\sim 10^{9}$ to  $\sim 10^{13}$~cm$^{-3}$. For comparison, the densities in X-ray coronal plasma can range $\sim 10^{8}$ to  $\sim 10^{13}$~cm$^{-3}$ depending on the activity of the star \citep{2004A&ARv..12...71G}.
For the flux rope magnetic field strength $B_\textsc{fr}$ we use values from \qtyrange{100}{1000}{\gauss}. Stellar flare observations estimate the magnetic field in the flare region to range from \qtyrange{30}{400}{\gauss}~\citep{2017ApJ...851...91N} and \citet{2019ApJ...880...97L} argue that starsport/stellar active regions could have field strengths of ~250 to 1400~G. The values we chose for $B_\textsc{fr}$ span the values found in these two studies. Another avenue that gives support to a stronger-than-solar small-scale field for AB Dor comes from observations of the total unsigned field of active stars. Typically, it is found that the large-scale field derived in spectropolarimetric observations (such as that presented in Fig.~\ref{fig:magnetogram}) encompasses only a fraction ($\sim 10\%$) of the total surface field (i.e., including the small and the large-scale fields, see, e.g., \citealt{2014MNRAS.441.2361V}). Given that the large-scale field alone of AB Dor is already a couple of orders of magnitude larger than solar, we speculate that this increase would be also found in the smaller-scale field structure. Of course, there is still a degeneracy between the filling factor of active regions and their typical field strengths---for example, we do not know whether a higher unsigned total field would result in large active regions with weaker field strengths or in smaller active regions with stronger magnetic fields.

The value of \(B_\textsc{fr}\) is used to set a current \(I_\textsc{fr} = 2 c R_\text{maj} B_\textsc{fr}/4\pi\) (\(c\) being the speed of light) that flows along the arc of the flux rope so that the magnetic field strength at the flux rope symmetry axis is \(B_\textsc{fr}\). In the literature, the flux rope parameter that is more commonly referred to is the free magnetic energy instead of the $B_\textsc{fr}$. The free magnetic energy is computed as \citep[see][]{1999A&A...351..707T,2003ApJ...588L..45R}
\begin{equation}\label{eq:freemag}
    E^{B}_\text{free} = \frac{1}{2} I_\textsc{fr}^2 L, \quad\text{with}
    \quad
    L= \frac{4\pi     }{c^2} \,
    \frac{\alpha}{\qty{360}{\degree}}R_\text{maj}  \,
    \left( \ln\frac{8R_\text{maj}}{R_\text{min}}-\frac54\right)
\end{equation}
where \(\alpha = \arccos{\big(D \big/ R_\text{maj}\big)}\) is the arch angle of the emerging flux rope and $L$ is its inductance. Since we only vary $B_\textsc{fr}$ and $M_\textsc{fr}$, we have that \(L=\qty{2.39e-10}{\square\second\per\centi\meter}\) in all our models. Likewise the parenthesised term in the expression for \(L\) is a geometrical factor with value \num{2.69}.
We note that $E^{B}_\text{free} \propto R_\text{maj}^3B_\textsc{fr}^2$ and that the flux rope mass does not affect the free energy.
Given our parameters for the flux rope, the free magnetic energy in the flux rope ranges from $4.5\times 10^{33}$ to $4.5\times 10^{35}$~erg. These values  match the range adopted in previous stellar CME studies  \citep{2018ApJ...862...93A, 2020ApJ...895...47A, 2022ApJ...924..115O}, with the lower bound values found in solar CME simulations \citep{2008ApJ...684.1448M, 2013ApJ...773...50J}. The higher end of our values are on the ballpark of energies observed in stellar superflares ($\gtrsim 10^{34}$ erg, e.g. \citealt{2013PASJ...65...49S, 2015EP&S...67...59M}). Finally, we also compute the poloidal flux $\phi_{\rm p}^\textsc{fr}$ of the flux rope, by integrating the unsigned magnetic field passing through a perpendicular cross section at the apex of the flux rope.  The first columns of Table~\ref{tab:main} show the parameters of the flux rope for each of the 21 CME simulations. Each combination of $M_\textsc{fr}$, $B_\textsc{fr}$, and flux rope location leads to a different CME evolution.

Three final parameters determine the location and orientation of the emerging flux rope on the stellar surface. In this study we either place the flux rope in the region of open magnetic field lines (green patch in Figure \ref{fig:magnetogram}) at a longitude of $-76^\circ$, a latitude of $55^\circ$ and an orientation of $-30^\circ$ with respect to east-west line, or at a region of closed magnetic field (magenta patch in Figure \ref{fig:magnetogram}) at a longitude of~$0^\circ$, latitude of~$65^\circ$ and an east-west orientation.
Note that we initialise our flux ropes at these high latitudes, because AB Dor shows sign of polar active regions  and a recent indication that it  hosts CMEs at polar regions (see Section~1).
The two positions were chosen to place the flux rope in the open and closed magnetic field regions of the star, respectively.
\begin{table*}
\caption{Initial flux rope properties (columns 2 through 6) and CME properties after one hour of evolution (columns 7 through 10). The ID of the simulation is shown in column one; he quantities inside the symbol are the flux rope mass and the magnetic field in the centre of the flux rope $B_\textsc{fr}$. A rounded (square) shape indicates that the flux rope is initiated in the open (closed) magnetic field region of the star.  The outline colour also gives a visual indication of $B_\textsc{fr}$.
The poloidal flux of the flux rope is $\phi_{p}^\textsc{fr}$ and its free magnetic energy is $E^{B}_\text{free}$. The following columns are: the maximum CME velocity $u_r$, the CME mass $M_\textsc{cme}$, the kinetic energy of the CME $E_\textsc{cme}^{k}$, and whether the flux rope resulted in an eruptive or confined CME.
The final two columns are the scaling relation-based values of $F^\text{fl}_\textsc{sxr}$ and $E_\text{fl}$ computed directly from $\phi_{p}^\textsc{fr}$ as described and discussed in Sect.~\ref{sec:cme_energetics}. }
\label{tab:main}
\renewcommand{\arraystretch}{1.3}
\begin{tabular}{cclcccccclcc} 
\hline ID                                       &  $M_\textsc{fr}$       &  Region  &  $B_\textsc{fr}$  &  $\phi_\text{p}^\textsc{fr}$  &  $E_\text{free}^{B}$     &  $u_r$                &  $M_\textsc{cme}$       &  $E^\text{k}_\textsc{cme}$    &  Behaviour & $F^\text{fl}_\textsc{sxr}(\phi_\text{p}^\textsc{fr})$ & $E_\text{fl}(\phi_\text{p}^\textsc{fr})$\\        
                                                &  [g]                   &          &  [G]              &  [Mx]                  &  [erg]                   &  [$\kms$]             &  [g]                    &  [erg]                 &   & [\si{\watt\per\square\meter}] & [erg]\\                          
\hline \SimCase{O}{200}{14}         &  $3.5 \times 10^{14}$  &  open    &  200              &  $5.3 \times 10^{21}$  &  $1.8 \times 10^{34}$    &  $2.2 \times 10^{2}$  &  $3.9 \times 10^{17}$   &  $9.1 \times 10^{31}$  &  Eruptive    & \num[round-mode=places,round-precision=1]{5.19e-05} &  $1.3 \times 10^{31}$ \\
\SimCase{O}{500}{14}                &  $3.5 \times 10^{14}$  &  open    &  500              &  $1.4 \times 10^{22}$  &  $1.1 \times 10^{35}$    &  $2.2 \times 10^{3}$  &  $1.6 \times 10^{16}$   &  $3.7 \times 10^{32}$  &  Eruptive    & \num[round-mode=places,round-precision=1]{2.61e-04} &  $4.6 \times 10^{31}$ \\                 
\SimCase{C}{500}{14}               &  $3.5 \times 10^{14}$  &  closed  &  500              &  $1.4 \times 10^{22}$  &  $1.1 \times 10^{35}$    &  -                    &  -                      &  -                     &  Confined    & \num[round-mode=places,round-precision=1]{2.69e-04} &  $4.7 \times 10^{31}$ \\      %
\SimCase{C}{750}{14}               &  $3.5 \times 10^{14}$  &  closed  &  750              &  $2.3 \times 10^{22}$  &  $2.5 \times 10^{35}$    &  -                    &  -                      &  -                     &  Confined    & \num[round-mode=places,round-precision=1]{6.66e-04} &  $9.6 \times 10^{31}$ \\     %
\SimCase{C}{1000}{14}              &  $3.5 \times 10^{14}$  &  closed  &  1000             &  $3.5 \times 10^{22}$  &  $4.5 \times 10^{35}$    &  $1.5 \times 10^{3}$  &  $1 \times 10^{16}$     &  $1.1 \times 10^{32}$  &  Eruptive    & \num[round-mode=places,round-precision=1]{1.35e-03} &  $1.7 \times 10^{32}$ \\                 
\hline \SimCase{O}{100}{16}         &  $3.5 \times 10^{16}$  &  open    &  100              &  $3.1 \times 10^{21}$  &  $4.5 \times 10^{33}$    &  -                    &  -                      &  -                     &  Confined    & \num[round-mode=places,round-precision=1]{2.01e-05} &  $6.0 \times 10^{30}$ \\       %
\SimCase{O}{200}{16}                &  $3.5 \times 10^{16}$  &  open    &  200              &  $5.3 \times 10^{21}$  &  $1.8 \times 10^{34}$    &  $1.4 \times 10^{3}$  &  $6.1 \times 10^{15}$   &  $1.2 \times 10^{34}$  &  Eruptive    & \num[round-mode=places,round-precision=1]{5.19e-05} &  $1.3 \times 10^{31}$ \\                 
\SimCase{O}{500}{16}                &  $3.5 \times 10^{16}$  &  open    &  500              &  $1.4 \times 10^{22}$  &  $1.1 \times 10^{35}$    &  $3.0 \times 10^{3}$  &  $2.1 \times 10^{16}$   &  $9.5 \times 10^{32}$  &  Eruptive    & \num[round-mode=places,round-precision=1]{2.61e-04} &  $4.6 \times 10^{31}$ \\                 
\SimCase{C}{500}{16}               &  $3.5 \times 10^{16}$  &  closed  &  500              &  $1.4 \times 10^{22}$  &  $1.1 \times 10^{35}$    &  -                    &  -                      &  -                     &  Confined    & \num[round-mode=places,round-precision=1]{2.69e-04} &  $4.7 \times 10^{31}$ \\           %
\SimCase{C}{650}{16}               &  $3.5 \times 10^{16}$  &  closed  &  650              &  $2.0 \times 10^{22}$  &  $1.9 \times 10^{35}$    &  -                    &  -                      &  -                     &  Confined    & \num[round-mode=places,round-precision=1]{5.23e-04} &  $7.9 \times 10^{31}$ \\           %
\SimCase{C}{900}{16}               &  $3.5 \times 10^{16}$  &  closed  &  900              &  $2.8 \times 10^{22}$  &  $3.6 \times 10^{35}$    &  -                    &  -                      &  -                     &  Confined    & \num[round-mode=places,round-precision=1]{9.06e-04} &  $1.2 \times 10^{32}$ \\   %
\hline \SimCase{O}{300}{17}         &  $3.5 \times 10^{17}$  &  open    &  300              &  $7.8 \times 10^{21}$  &  $4.0 \times 10^{34}$    &  $2.4 \times 10^{3}$  &  $2.4 \times 10^{18}$   &  $7.1 \times 10^{34}$  &  Eruptive    & \num[round-mode=places,round-precision=1]{1.02e-04} &  $2.2 \times 10^{31}$ \\                 
\SimCase{C}{900}{17}               &  $3.5 \times 10^{17}$  &  closed  &  900              &  $2.8 \times 10^{22}$  &  $3.6 \times 10^{35}$    &  $1.5 \times 10^{3}$  &  $5.4 \times 10^{16}$   &  $5.7 \times 10^{32}$  &  Eruptive    & \num[round-mode=places,round-precision=1]{9.06e-04} &  $1.2 \times 10^{32}$ \\                 
\hline \SimCase{O}{100}{18}         &  $3.5 \times 10^{18}$  &  open    &  100              &  $3.1 \times 10^{21}$  &  $4.5 \times 10^{33}$    &  $6.0 \times 10^{1}$  &  $3.5 \times 10^{17}$   &  $8.4 \times 10^{30}$  &  Confined    & \num[round-mode=places,round-precision=1]{2.01e-05} &  $6.0 \times 10^{30}$ \\                 
\SimCase{O}{300}{18}                &  $3.5 \times 10^{18}$  &  open    &  300              &  $7.8 \times 10^{21}$  &  $4.0 \times 10^{34}$    &  $9.9 \times 10^{2}$  &  $9.2 \times 10^{17}$   &  $4.5 \times 10^{33}$  &  Eruptive    & \num[round-mode=places,round-precision=1]{1.02e-04} &  $2.2 \times 10^{31}$ \\                 
\SimCase{O}{1000}{18}               &  $3.5 \times 10^{18}$  &  open    &  1000             &  $3.5 \times 10^{22}$  &  $4.5 \times 10^{35}$    &  $3.8 \times 10^{3}$  &  $4.1 \times 10^{18}$   &  $2.9 \times 10^{35}$  &  Eruptive    & \num[round-mode=places,round-precision=1]{1.36e-03} &  $1.7 \times 10^{32}$ \\                 
\SimCase{C}{300}{18}               &  $3.5 \times 10^{18}$  &  closed  &  300              &  $7.7 \times 10^{21}$  &  $4.0 \times 10^{34}$    &  $3.9 \times 10^{0}$  &  $1.5 \times 10^{17}$   &  $1.2 \times 10^{28}$  &  Confined    & \num[round-mode=places,round-precision=1]{9.79e-05} &  $2.1 \times 10^{31}$ \\                 
\SimCase{C}{500}{18}               &  $3.5 \times 10^{18}$  &  closed  &  500              &  $1.4 \times 10^{22}$  &  $1.1 \times 10^{35}$    &  $1.0 \times 10^{1}$  &  $1.3 \times 10^{17}$   &  $6.3 \times 10^{28}$  &  Confined    & \num[round-mode=places,round-precision=1]{2.69e-04} &  $4.7 \times 10^{31}$ \\                 
\SimCase{C}{650}{18}               &  $3.5 \times 10^{18}$  &  closed  &  650              &  $2.0 \times 10^{22}$  &  $1.9 \times 10^{35}$    &  $8.3 \times 10^{1}$  &  $5.4 \times 10^{17}$   &  $1.9 \times 10^{31}$  &  Confined    & \num[round-mode=places,round-precision=1]{5.23e-04} &  $7.9 \times 10^{31}$ \\                 
\SimCase{C}{800}{18}               &  $3.5 \times 10^{18}$  &  closed  &  800              &  $2.5 \times 10^{22}$  &  $2.9 \times 10^{35}$    &  $8.3 \times 10^{2}$  &  $6.1 \times 10^{17}$   &  $2.1 \times 10^{33}$  &  Eruptive    & \num[round-mode=places,round-precision=1]{7.43e-04} &  $1.0 \times 10^{32}$ \\                 
\SimCase{C}{1000}{18}              &  $3.5 \times 10^{18}$  &  closed  &  1000             &  $3.5 \times 10^{22}$  &  $4.5 \times 10^{35}$    &  $1.7 \times 10^{3}$  &  $7.6 \times 10^{17}$   &  $1.1 \times 10^{34}$  &  Eruptive    & \num[round-mode=places,round-precision=1]{1.35e-03} &  $1.7 \times 10^{32}$ \\                 
\hline 
\end{tabular} 

\includegraphics{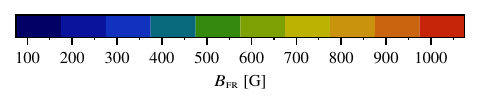}
\end{table*}

Table \ref{tab:main} shows the initial parameters of the flux ropes alongside resulting CME characteristics (presented in Section \ref{sec.CME}).
To distinguish between our various models, each model has a unique identifier symbol. The numbers inside each symbol show the flux rope mass in grams, and the flux rope field strength in Gauss. The colour also indicates the magnetic field strength of the flux rope according to the colourbar shown in Table \ref{tab:main}. The shape of the symbol indicates whether the flux rope is initiated in a closed (square) or open (rounded) magnetic field region. For instance, the symbol \SimCase{C}{1000}{18} means that the flux rope has a
    mass of $3.5\times 10^{18}\,\text{g}$,
    a magnetic field of 1000~G, and is located at a region where the stellar magnetic field is closed (square). In the same way, the symbol \SimCase{O}{200}{16} represents a flux rope with
    mass $3.5\times 10^{16}\,\text{g}$,
    a magnetic field of 200~G, located in an open field region (rounded).

\section{Physical properties of our simulated CMEs}\label{sec.CME}
\subsection{Overview of our simulated CMEs}
\begin{figure*}
    \centering
    \includegraphics[width =\textwidth]{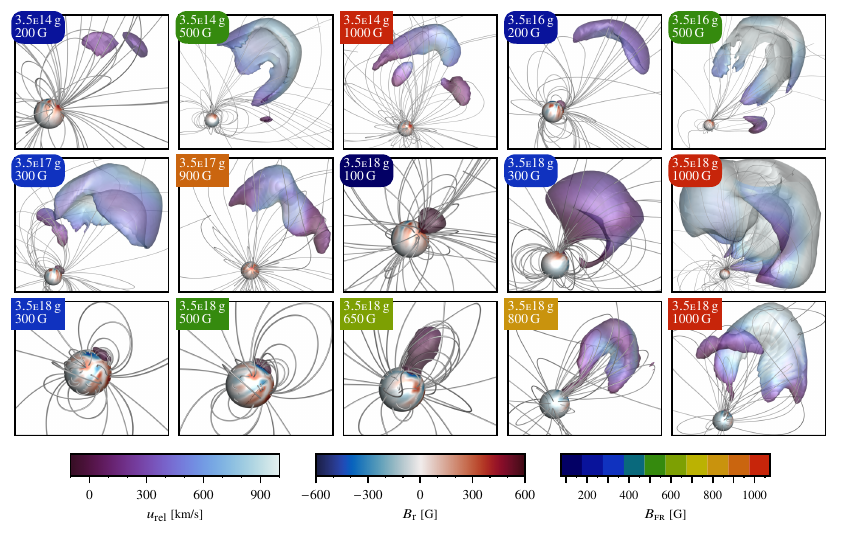}
    \caption{Snapshots after one hour of evolution. The magnetic field of the initial flux rope $B_\textsc{fr}$ (symbol colour) increases from top left to bottom right for each initial flux rope mass $M_\textsc{fr}$ (cf.~symbol text). Flux ropes located in open field lines (rounded symbols) of the background wind evolve faster than those located in closed (square symbols) field line regions, for the same $B_\textsc{fr}$ and $M_\textsc{fr}$. Isosurfaces are defined using a plasma density ratio \(n/n_\text{ss}\) of four and are colour coded by their relative radial velocity compared to the local stellar wind value.
    The confined events with
    \(M_\textsc{fr}<\qty{e17}{\gram}\)
    are not shown in this figure as there was no overdense \(n/n_\text{ss}\geq 4\) region remaining after one hour of evolution.}\label{fig:CME_runs}
\end{figure*}

We injected 21 different flux rope configurations into our background stellar wind model and let the eruption evolve. Figure \ref{fig:CME_runs} shows the 3D images of 15 simulated CMEs at a snapshot of 1 hour evolution. The isosurface, used to track the CME evolution, shows where the  ratio between the wind particle density $n$ at a given simulation time and the steady-state stellar wind $n_\text{ss}$ is $n/n_\text{ss} = 4$. The isosurface colour represents the relative speed of the CME compared to the background wind: $u_\text{rel} = u_{r} - u_{r\! , \, \text{ss}}$ and the inner boundary is coloured by the surface magnetic field strength. %
 Six models are not shown in Figure~\ref{fig:CME_runs}, as no overdense region remains after one hour.

Several factors seem to influence the resulting CME morphology. Because the energy released from the erupting flux rope must overcome the overlying magnetic tension and open the field lines to escape, the location of the flux rope with respect to the background stellar wind magnetic field changes the amount of magnetic energy needed for a CME to escape~\cite[e.g.][]{2018ApJ...862...93A}. The magnetic tension of the stellar magnetic field is larger in regions of closed field lines, and hence a higher flux rope energy is required to break up the overlying field compared to flux ropes injected in open field regions. This can be seen for example in model \SimCase{O}{1000}{18} in Figure \ref{fig:CME_runs}, where we find the most extended CME at 1~h of evolution. This case has the most energetic flux rope configuration (largest mass and largest magnetic field) and is located in the region of open field lines. Model \SimCase{C}{1000}{18}  shows a flux rope with identical parameters, but now located in the closed-field line region. While this CME is also sufficiently energetic to break up the magnetic tension, it propagates at a slower speed and, hence, does not reach the same distance as the case in the open field line region \SimCase{O}{1000}{18}. It should be noted that it is not only the initial overlying field strength that influence the propagation of the CME but also the rate at which the overlying field is decaying with stellar height \citep{deng2017roles}.

Figure~\ref{fig:CME_runs} also show cases with nearly no CME evolution like model \SimCase{C}{300}{18}, where the flux rope energy is  not sufficient to overcome the overlying magnetic tension. This model has a weaker flux rope field strength of 300~G. This event is classified as confined and has an average speed of $-29\kms$ (i.e., material falling back to the star) after 1~h of evolution (more about CME confinement is discussed in Section \ref{sec:cme_confinement}). Next to this model, we have the event \SimCase{O}{300}{18}, generated with the same flux rope configuration but located in the open field region. Despite their similar flux rope characteristics, this CME has a mean average speed of $716\kms$. The energy in this case is sufficient for the CME to break through the overlying field.
CME \SimCase{O}{300}{18} escapes and acquires a mass of around $9.2 \times 10^{17}\,$g, while the counterpart model \SimCase{C}{300}{18} has a mass of about $1.5 \times 10^{17}\,$g, but remained confined. We continue the discussion about the mass, velocity and energetics of the CMEs in Section \ref{sec:cme_energetics}.

As the CME propagates through the stellar wind, it changes the local properties of the wind. Our model \SimCase{C}{1000}{18}, for example, reaches a radial velocity at its apex of $\sim 1700\,\kms$. The local plasma temperature during the acceleration process of the CME can reach up to tens of millions of K close to the shock created in front of the expanding CMEs apex, while temperatures behind the CME (typically in regions where plasma is evacuated) can be a few orders of magnitude colder. During this time, eruptive and energetic CMEs are extremely overdense and can, in some regions, yield plasma density ratios of a few thousand times that of the local stellar wind.

CME  \SimCase{C}{1000}{14} is another notable case, which fragmented over time and generated one main and two smaller CME regions (similar effects has been shown in e.g., \citealp{2022ApJ...928..147A}).
The local wind in front of the CME is heated and accelerated to velocities that are about a factor of 3 of the background wind.
This CME creates a region with speeds exceeding the local Alfvén speed and produces a shock.
Compared to the main CME fragment, the smaller child fragments in model \SimCase{C}{1000}{14} form regions of slower wind speeds behind them, with speeds down to about half of the background (original) wind speed. The smaller fragmented pieces are not fast enough to generate a shock and are still attached to closed magnetic field lines.

\subsection{Confined versus eruptive CMEs}\label{sec:cme_confinement}
\begin{figure}
    \centering
    \includegraphics[width = .48\textwidth]{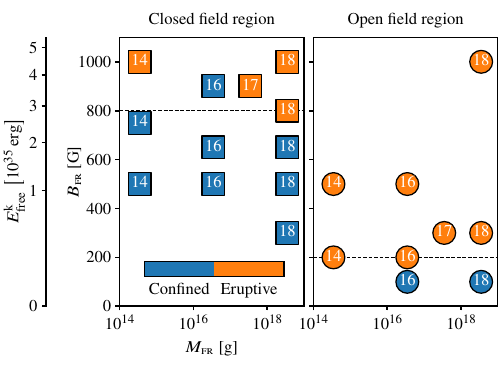}
    \caption{Summary of the eruptive and confined events in our 21 simulated CMEs. The $x$ and $y$ axes represent the assumed field strength and mass of the flux rope, before its eruption, respectively. Of the 21 simulated events, 10 are classified as confined.
    In each panel a dashed line has been added to indicate the  lowest \(B_\textsc{fr}\) value of an eruptive event. The left panel shows the events initiated in the closed-field region, while the right panel shows the events initiated in the open field region.
    The free magnetic energy from eq.~\eqref{eq:freemag} and Table~\ref{tab:main} shown in the leftmost axis. Note that $E_\text{k}^\text{free} \propto B_\textsc{fr}^2$.
    }
    \label{fig:Simulation_space}
\end{figure}

Figure \ref{fig:Simulation_space} summarises the eruptive/confined nature of all the 21 events we simulated. We define a CME to be eruptive by manually inspecting the CME morphology after 1~h of evolution and confirming that the CME has a positive average velocity at this time. %
The left (right) panel shows the events that were initiated in the region where the stellar magnetic field has closed (open) configurations. The events marked in orange are the ones that erupted, while the non-eruptive events are shown in blue. In total, we have 10/21 confined events.

Overall, our ten confined events tend to be located in regions of closed field lines (their symbols are squares), indicating that the input flux rope energy is too weak to break the overlying field tension. The only two confined events located in open field lines, \SimCase{O}{100}{16} and \SimCase{O}{100}{18}, have a weak flux rope magnetic field $B_\textsc{fr}=100\,\text{G}$. In other words, for a CME to erupt, its initial flux rope requires a higher magnetic field if it is located in a region of closed field lines compared to events taking place in open field line regions. In our simulations, no models with initial flux rope magnetic fields $B_\textsc{fr} \lesssim 800$~G were eruptive for closed-field CMEs. For open-field CME models, the threshold was $B_\textsc{fr} \lesssim 200$~G (horizontal lines Figure \ref{fig:Simulation_space}).

Additionally, some eruptive CMEs leave behind a dense region of plasma. This can be seen in models \SimCase{O}{200}{16} and \SimCase{O}{300}{17}, where we see a blob of plasma loitering at the inner boundary. In this case, a fragment of the erupting CMEs mass is falling back to the stellar surface because it is unable to overcome the gravitational potential of the star. The plasma trapped close to the stellar surface is three orders of magnitudes denser that the main front of the CME in  these models. We classify these events as eruptive, because the bulk of the CME mass has escaped.

\subsection{CME energetics} \label{sec:cme_energetics}
\begin{figure}
    \centering
    \includegraphics[clip,trim={0 0.60cm 0 0.2cm},width = .46\textwidth]{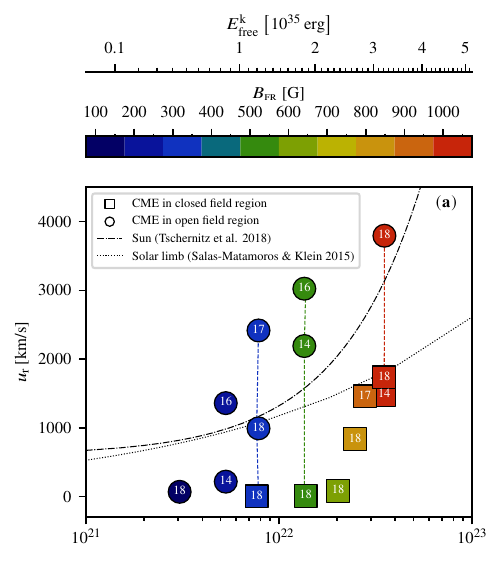}\\
    \includegraphics[clip,trim={0 0.60cm 0 0.2cm},width = .46\textwidth]{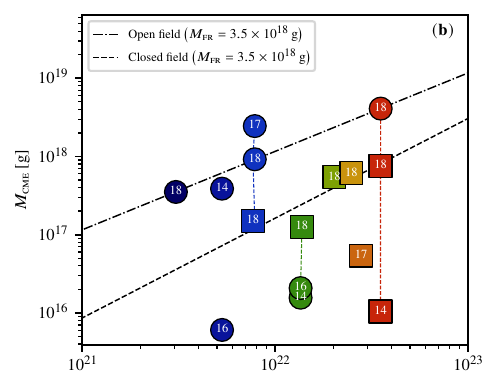}\\
    \includegraphics[clip,trim={0 0.25cm 0 0.2cm},width=0.46\textwidth]{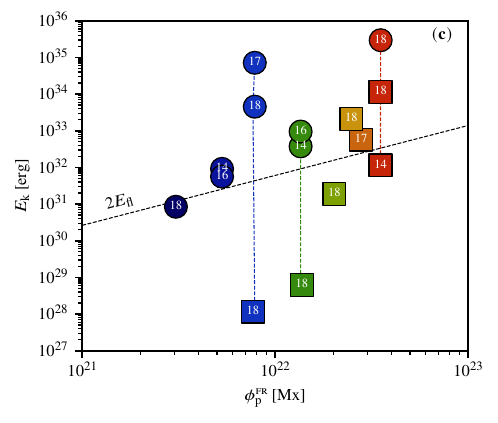}
    \caption{Maximum speed (a), mass (b) and kinetic energies (c) of the simulated CMEs after 1~h of evolution as a function of poloidal magnetic flux of the flux rope. Symbols are as presented in Table \ref{tab:main}. In panel (a), we show the two relations derived from solar CMEs as dotted and dash-dotted lines. In panel (b), we fit power-laws to the cases with flux rope masses $M_\textsc{fr} = 3.5 \times 10^{18}$~g.
    In panel (c), the dashed line stands for the energy threshold \(E_\text{k} = 2E_\text{fl}\), which is a good proxy separating confined or eruptive events.
    }\label{fig:energetics}
\end{figure}

To compute the energetics of the simulated CMEs, we use their radial speeds and masses. The velocities we present in this work are the maximum speed of the material enclosed in the volume of the CME, which is defined by a density ratio $n/n_\text{ss} \geq 4$ \citep{2022ApJ...924..115O}. The masses in our models are calculated by integrating the density within this volume. Note that other studies apply different values for the ratio when determining the CME region  \citep[e.g.,][ use a factor of 3 instead]{2020ApJ...895...47A}. By decreasing this ratio, a larger CME volume is produced, with thus a higher mass.
(We calculated the CME masses using either a density ratio of 4 or 3, but found results within the same order of magnitude.) The kinetic energy of each CME is then calculated from the CME mass and velocity. Figure \ref{fig:energetics} shows the maximum speeds, masses and kinetic energies of each CME as a function of the poloidal magnetic flux of the flux rope $\phi_\text{p}^\textsc{fr}$. We calculate this by integrating the unsigned magnetic field passing through a perpendicular cross section at the apex of the flux rope.  It is interesting to note that the poloidal magnetic flux of the flux rope has been used as a proxy for X-ray flare energy in stellar CMEs studies \citep{2018ApJ...862...93A}.

We see in Figure \ref{fig:energetics}a  that, after 1~h of evolution, the maximum speeds of the eruptive events range from 220 to $3700\,\kms$, while the lowest velocity points refer to confined events. Note that in this figure, we do not include the six confined events that did not propagate outward after 1~h (\SimCase{C}{500}{14}, \SimCase{C}{750}{14}, \SimCase{O}{100}{16}, \SimCase{C}{500}{16}, \SimCase{C}{650}{16}, \SimCase{C}{900}{16}). The maximum CME speed we found is around five times higher than the average background wind velocity ($u_\text{r, ss} \simeq 720\,$km\,s$^{-1}$) at the 1~h mark. For the eruptive models, CMEs occurring in the closed regions (square symbols) have lower radial velocities than the ones occurring in open-field regions (circle symbols), for the same flux rope conditions (e.g., models \SimCase{C}{1000}{18} and \SimCase{O}{1000}{18}). This is due to the stronger magnetic tension exerted on the CME material by the overlying closed magnetic field lines.

For comparison, we also plot in Figure \ref{fig:energetics}a the relations derived from solar wind studies from \citet[dashed-dotted line]{tschernitz2018reconnection} and \citet[dotted line]{2015SoPh..290.1337S}, which were re-written as functions of $\phi_p^\textsc{fr}$ in the work of \citet{2018ApJ...862...93A}. CMEs in open regions agree better with the solar relation from \citet{tschernitz2018reconnection}, while closed region CMEs show  lower speeds than the solar trends.
Solar CME speeds range from a few hundred up to a few thousand $\kms$ (\citealp{2005JGRA..11012S05Y}).

From Figure \ref{fig:energetics}b, we find that both the mass and poloidal flux of the flux rope govern the mass of the ejected CME, agreeing with the findings of \citet{2018ApJ...862...93A}.
For flux ropes of similar masses, we find that increasing the poloidal flux (and therefore the hypothetically assumed flare energy) in general lead to larger CME masses after 1~h of evolution. This is in agreement with observations of solar and stellar CMEs \citep[e.g.][]{2011SoPh..268..195A, 2013ApJ...764..170D, 2016ApJ...833L...8T, 2019ApJ...877..105M}, where CME mass and speed are found to increase with increasing X-ray flare energy, typically used as a proxy for poloidal flux \citep{2018ApJ...862...93A}.
Figure \ref{fig:energetics}b shows two power-law fits to the cases with flux rope masses of $M_\textsc{fr} = 3.5 \times 10^{18}$~g, where we found that
$M_\textsc{cme} [\textrm{g}] = 10^{-4.001}(\phi_\text{p} ^\textsc{fr} [\textrm{Mx}])^{1.003}$ and
$M_\textsc{cme} [\textrm{g}] = 10^{-10.85}(\phi_\text{p} ^\textsc{fr} [\textrm{Mx}])^{1.276}$ for the open (dash-dotted line) and closed (dashed line) field models, respectively. There is a faster increase in the CME mass with poloidal flux for the events located in closed field line regions, even though their CME masses are smaller than the ones placed in open field line regions (see below).
As we discuss further in Section 4, we caution the reader that in our parametric study, we do not vary the geometry of the flux rope, which would also affect the amount of free energy available for driving the CME. Therefore, the relations shown in Figure 6 are limited to this assumption of our model.

Although there is an increase of CME mass with poloidal flux, we also see that for the same poloidal flux and varying range of flux rope masses, our models predict a range of CME masses. In general,  the most massive flux ropes tend to generate CMEs that are more massive. Models with flux rope masses of $3.5 \times 10^{18}$~g evolve into CMEs with masses ranging from $1.3 \times 10^{17}$~g to $4.1 \times 10^{18}$~g after 1\,h of evolution. (For reference, the 1859 Carrington solar CME event had en estimated mass of $8 \times 10^{16}$~g,  \citealt{2013JSWSC...3A..31C}.)
We found that CMEs with initially smaller flux rope masses typically collect material while propagating through the stellar wind. This results in an increase in the evolved CME mass by up to two orders of magnitude (see models \SimCase{O}{500}{14} and \SimCase{C}{1000}{14}, which have CME masses above the initial flux rope masses). Conversely, CMEs with initially higher flux rope masses are prone to losing mass, either since parts of the CME are confined and fall back onto the stellar surface, or because the CME becomes more attenuated as it propagates outwards (the mass is then decreased since less of the plasma satisfy the condition that $n/n_\text{ss}\geq4$). The two models deviating from this behaviour are \SimCase{O}{300}{17} and \SimCase{O}{1000}{18}, where the evolved CME mass is larger than the initial flux rope mass.

Models with similar initial values of $M_\textsc{fr}$ and $B_\textsc{fr}$, but initiated in different (open/closed) regions, exhibit different CME masses  after they have evolved.
In open regions, a larger fraction of the initial flux rope mass of the CME becomes energetic enough to escape, which results in a higher CME masses (compare, e.g., \SimCase{O}{300}{18} with \SimCase{C}{300}{18}, or \SimCase{O}{1000}{18} with \SimCase{C}{1000}{18}). This is also seen in the fits provided in panel b, where the fit for the open field line lies above the closed-field one.

Figure \ref{fig:energetics}c shows the kinetic energies of our simulated CMEs, calculated as $\frac12 M_\textsc{cme} u_r^2$. In this figure, we only plot models with  positive velocities at 1~h. We also show the eruptive/confined energy threshold (dashed line), where a CME is defined as confined if  $E_\text{k} < 2E_\text{fl}$, and eruptive otherwise \citep{2018ApJ...862...93A}. The four lowest points in Figure \ref{fig:energetics}c correspond to confined events, while the other points correspond to eruptive events. Therefore, our models agree well with the energy threshold (dashed line). %
Some confined CMEs (in particular \SimCase{C}{300}{18} and \SimCase{C}{500}{18}) have energies even below $10^{29}\,$erg. The kinetic energies of eruptive CMEs range  from about $10^{31}$ to $10^{35}\,$erg.

\citet{2018ApJ...862...93A} presented a way to connect the poloidal flux, derived from simulation parameters, to the observed GOES peak soft X-ray fluxes of solar flares. Following this approach, we use the relations between flare energy ($E_{\rm fl}$), soft X-ray fluxes ($F_\textsc{sxr}^\text{fl}$) and poloidal fluxes ($\phi_{\rm p}^\textsc{fr}$), namely,
\begin{equation*}
    E_{\rm fl} = 10^{34.49} [F_\textsc{sxr}^\text{fl}]^{0.79}
\end{equation*}
and
\begin{equation*}
\phi_{\rm p}^\textsc{fr} =10^{24.21}[F_\textsc{sxr}^\text{fl}]^{0.580}
\end{equation*}
\citep{tschernitz2018reconnection, 2018ApJ...862...93A}, to compute the flare energy as
\begin{equation*}
E_{\rm fl} =10^{ 1.5143} [\phi_{\rm p}^\textsc{fr}]^{1.3621}.
\end{equation*}
Here the units of \(\phi_{\rm p}^\textsc{fr}\), \(F^\text{fl}_\textsc{sxr}\) and \(E_\text{fl}\) are \si{Mx}, \si{\watt\per\square\meter} and \si{erg}, respectively. With these relations, the \(\phi_\text{p}^\textsc{fr}\) in our CME models correspond to flare fluxes \(F_\textsc{sxr}^\text{fl}\) ranging from \qtyrange{2.0e-5}{1.4e-3}{\watt\per\square\meter}
and flare energies \(E_\text{fl}\) from \qtyrange{6.0e30}{1.7e32}{erg} (see Table~\ref{tab:main}).
The most energetic CME, model \SimCase{O}{1000}{18}, corresponds to a flare flux of $F_\textsc{sxr}^\textrm{fl} \approx 1.4 \times 10^{-3}\,$W\,m$^{-2}$, which would have been classified as a GOES class~X10 CME.
This CME acquires a velocity of $\sim 3800\,\kms$ with a total mass of $\sim 4.1 \times 10^{18}\,$g and a kinetic energy of $2.9 \times 10^{35}\,$erg at 1~h of evolution. For reference, most solar CMEs have observed masses in ranges of $10^{14}$ -- $10^{16}$~g with velocities reaching up to around $1000\,\kms$ and kinetic energies in the interval $10^{28}$ -- $10^{30}\,$erg \citep{2010ASSP...19..289G, webb2012coronal}. Note that a small part of the CME mass in this model reached the edge of the model domain at 1~h of evolution and that its energies therefore can be slightly underestimated.

Finally, the resulting CMEs depend on whether the flux rope is injected in regions of open or closed field lines. Model \SimCase{C}{1000}{18} is initiated with the same flux rope properties as \SimCase{O}{1000}{18}, but in a closed-field region. The overlying field is able to reduce the CME energy to only $\sim 4\,$\% of the energy of the equivalent model initiated in the open region. We can also compare the two similar models \SimCase{O}{300}{18} and \SimCase{C}{300}{18}. These models also have the same initial flux rope setup, differing only in terms of the overlying magnetic field. The closed-field model has its energy reduced by about six orders of magnitude, and remains confined.  In the cases of models \SimCase{O}{1000}{18} and \SimCase{C}{1000}{18}, both models have enough input energy to escape the overlying field and still have sufficient energy left after 1~h of evolution. %
These results demonstrate that not only the energetics of the flux rope but also the property of the stellar wind background field, are key properties in shaping the energetics and the evolution of the CME.

\section{Discussion and Conclusion}\label{sec:discussion_and_conclusion}

It has been expected that active stars, which flare more frequently than the Sun, would have more frequent ejections of CMEs \citep{2012ApJ...760....9A, 2013ApJ...764..170D, 2015ApJ...809...79O}. Yet, it has been very challenging to detect stellar CMEs \citep{2022arXiv221105506N}. Because of this, it has been suggested that CMEs in active stars might be more frequently confined \citep{2018ApJ...862...93A, 2022MNRAS.509.5075S}. One reason for this is that the strong overlying surface magnetic fields observed in these stars would not allow the material to break open the overlying magnetic tension, thus confining the eruption of a CME. However, recently, \citet{2021NatAs...5..697V} reported the detection of coronal dimmings, which are indicative of the presence of eruptive CMEs. One of these stars, the young and active solar-type star AB Dor, presented a high-latitude CME \citep{2021NatAs...5..697V}.

In this work, we investigate this by conducting a parametric study with 21 simulations of CME events on AB Dor. AB Dor shows polar activity in the form of polar spots, so all our modelled CMEs were injected in high-latitude  ($\lambda\sim\qty{60}{\deg}$) regions. Firstly, we modelled the steady-state  wind of AB Dor, using a reconstructed magnetic map based on spectropolarimetric observations made in December, 2002~\citep{2007MNRAS.377.1488H}. Secondly, flux ropes with different properties were initiated  and allowed to evolve through  the  steady background wind. We used the model of \citet{1999A&A...351..707T} to describe the flux rope. In our parametric study, we only assumed one particular geometry for the flux ropes, but the central magnetic field strength of the flux rope, the mass of the flux rope and also the region where the flux rope was placed varied in each run. The flux rope was either placed in a region with an overlying closed magnetic field or in a region with an overlying open magnetic field. We then performed a series of analyses to the CME events that were generated. Our findings can be summarised as follows:

About half of our modelled events resulted in eruptive CMEs, with the other half leading to confined CMEs.  CMEs initiated in open regions were more likely to be eruptive compared to those initiated in closed regions. In our simulations, no models with initial flux rope magnetic fields under $\sim 800$~G were eruptive for closed-field CMEs. For open-field CME models, the limit was around 200 G (Figure \ref{fig:Simulation_space}).

The initial parameters and the position of the flux rope govern the evolved morphology and energetics of the CME. Models with larger input poloidal fluxes (i.e., flux ropes with higher magnetic fields) produce, in general, CMEs with higher velocities, higher masses and thus higher kinetic energies. We also find that the initial mass of the flux rope govern the final mass and therefore kinetic energies of the produced CMEs, with flux ropes that are more massive tending to generate more massive CMEs (but note also the large spread seen in Figure \ref{fig:energetics}).

Closed-region CMEs require more initial energy in their input flux ropes to break open the overlying magnetic field line and overcome the magnetic tension. Therefore, the energetics of the CME after 1~h of evolution is strongly dependent on the overlying field geometry, where eruptive CMEs produced in an open field region are generally faster and more massive (i.e., higher kinetic energies) compared to eruptive CMEs produced in the closed-field region.
Our eruptive closed-field line CMEs attain lower velocities compared to what we ordinarily observe for solar CMEs after 1h of evolution. %
Conversely, the majority of eruptive open-field line CMEs are faster than solar CMEs after 1h of evolution.

We also computed  observable quantities, such as flare energy and flare fluxes, from model parameters (poloidal magnetic flux of the flux rope), following the approach used in   \citet{2018ApJ...862...93A}. In our models, for a given poloidal flux with different flux rope masses, we have only one flare flux (or one flare energy) estimate, but multiple CME properties. This implies that we cannot uniquely characterise the energetics of CMEs only from observationally-derived flare quantities.

 The free magnetic energy in our flux rope depends on both the size and the magnetic field of the flux rope ($E^{B}_\text{free} \propto R_\text{maj}^3 B_\textsc{fr}^2$, see Equation 1). Given that in our parametric study, we only used one particular geometry for the flux rope,  all the variation assumed in the free magnetic energy only comes from the variation on $B_\textsc{fr}$. This is a limitation of our parametric study, as the length scales of flaring loops can vary by several orders of magnitude \citep{1999ApJ...526L..49S,2002ApJ...577..422S, 2017ApJ...851...91N}. In the future, it would interesting to expand our parametric study to also include different sizes of flux ropes. For example, there are combinations of rope geometries and field strengths that would generate the same free magnetic energy and it would be interesting to investigate how this combination of parameters would affect the CME energetics.

Our work explains how eruptive CMEs such as those observed by \citet{2021NatAs...5..697V} may occur even in stars such as AB~Dor whose strong overlying magnetic fields should lead to more confined CMEs~\citep{2018ApJ...862...93A}.
We propose that, in stars with polar active regions, where the overlying magnetic field lines can become open,  CMEs can more easily escape the stellar surface and propagate into the stellar wind. It is also interesting to note that in our models, the most energetic CMEs expand considerably, including in the direction towards the equatorial plane, where potentially planets would be orbiting. Therefore, even though the CME originates in stellar polar regions, they would still affect the space weather environment around planets orbiting in the equatorial plane.
As the observational epochs of the dimming event and the surface magnetic map observations are not contemporaneous (1994 and 2002, respectively), it is possible that the stellar surface magnetic field has  evolved between the two epochs. Therefore, simultaneous observations probing potential CMEs and the stellar magnetic field, such as those of~\citet{2024ApJ...961...23N}, are of great value.

\section*{Acknowledgements}
We thank the anonymous referee for their constructive and thoughtful comments, which helped improve the quality of our manuscript. This project has received funding from the European Research Council (ERC) under the European Union's Horizon 2020 research and innovation programme (grant agreement No 817540, ASTROFLOW).
This work used the Dutch national e-infrastructure with the support of the SURF Cooperative using grant nos. EINF-2218 and EINF-5173.
This research has made use of NASA's \href{https://ui.adsabs.harvard.edu/}{Astrophysics Data System}.
This work was carried out using the SWMF tools developed at The University of Michigan \href{https://spaceweather.engin.umich.edu/the-center-for-space-environment-modelling-csem/}{Center for Space Environment Modelling (CSEM)} and made available through the NASA \href{https://ccmc.gsfc.nasa.gov/}{Community Coordinated Modelling Center (CCMC)}.
This work has made use of the following additional numerical software and visualisation software:
      NumPy~\citep{2011CSE....13b..22V},
      SciPy~\citep{2020SciPy-NMeth},
      and
     Matplotlib~\citep{2007CSE.....9...90H}.

\section*{Data Availability}
The data underlying this article will be shared on reasonable request to the corresponding author.

\bibliographystyle{mnras}
\bibliography{example} %

\begin{thebibliography}{}
\makeatletter
\relax
\def\mn@urlcharsother{\let\do\@makeother \do\$\do\&\do\#\do\^\do\_\do\%\do\~}
\def\mn@doi{\begingroup\mn@urlcharsother \@ifnextchar [ {\mn@doi@}
  {\mn@doi@[]}}
\def\mn@doi@[#1]#2{\def\@tempa{#1}\ifx\@tempa\@empty \href
  {http://dx.doi.org/#2} {doi:#2}\else \href {http://dx.doi.org/#2} {#1}\fi
  \endgroup}
\def\mn@eprint#1#2{\mn@eprint@#1:#2::\@nil}
\def\mn@eprint@arXiv#1{\href {http://arxiv.org/abs/#1} {{\tt arXiv:#1}}}
\def\mn@eprint@dblp#1{\href {http://dblp.uni-trier.de/rec/bibtex/#1.xml}
  {dblp:#1}}
\def\mn@eprint@#1:#2:#3:#4\@nil{\def\@tempa {#1}\def\@tempb {#2}\def\@tempc
  {#3}\ifx \@tempc \@empty \let \@tempc \@tempb \let \@tempb \@tempa \fi \ifx
  \@tempb \@empty \def\@tempb {arXiv}\fi \@ifundefined
  {mn@eprint@\@tempb}{\@tempb:\@tempc}{\expandafter \expandafter \csname
  mn@eprint@\@tempb\endcsname \expandafter{\@tempc}}}

\bibitem[\protect\citeauthoryear{{Aarnio}, {Stassun}, {Hughes}  \&
  {McGregor}}{{Aarnio} et~al.}{2011}]{2011SoPh..268..195A}
{Aarnio} A.~N.,  {Stassun} K.~G.,  {Hughes} W.~J.,   {McGregor} S.~L.,  2011,
  \mn@doi [\solphys] {10.1007/s11207-010-9672-7}, \href
  {https://ui.adsabs.harvard.edu/abs/2011SoPh..268..195A} {268, 195}

\bibitem[\protect\citeauthoryear{{Aarnio}, {Matt}  \& {Stassun}}{{Aarnio}
  et~al.}{2012}]{2012ApJ...760....9A}
{Aarnio} A.~N.,  {Matt} S.~P.,   {Stassun} K.~G.,  2012, \mn@doi [\apj]
  {10.1088/0004-637X/760/1/9}, \href
  {https://ui.adsabs.harvard.edu/abs/2012ApJ...760....9A} {760, 9}

\bibitem[\protect\citeauthoryear{{Alvarado-G{\'o}mez}, {Drake}, {Cohen},
  {Moschou}  \& {Garraffo}}{{Alvarado-G{\'o}mez}
  et~al.}{2018}]{2018ApJ...862...93A}
{Alvarado-G{\'o}mez} J.~D.,  {Drake} J.~J.,  {Cohen} O.,  {Moschou} S.~P.,
  {Garraffo} C.,  2018, \mn@doi [\apj] {10.3847/1538-4357/aacb7f}, \href
  {https://ui.adsabs.harvard.edu/abs/2018ApJ...862...93A} {862, 93}

\bibitem[\protect\citeauthoryear{{Alvarado-G{\'o}mez}
  et~al.,}{{Alvarado-G{\'o}mez} et~al.}{2020}]{2020ApJ...895...47A}
{Alvarado-G{\'o}mez} J.~D.,  et~al., 2020, \mn@doi [\apj]
  {10.3847/1538-4357/ab88a3}, \href
  {https://ui.adsabs.harvard.edu/abs/2020ApJ...895...47A} {895, 47}

\bibitem[\protect\citeauthoryear{{Alvarado-G{\'o}mez}
  et~al.,}{{Alvarado-G{\'o}mez} et~al.}{2022}]{2022ApJ...928..147A}
{Alvarado-G{\'o}mez} J.~D.,  et~al., 2022, \mn@doi [\apj]
  {10.3847/1538-4357/ac54b8}, \href
  {https://ui.adsabs.harvard.edu/abs/2022ApJ...928..147A} {928, 147}

\bibitem[\protect\citeauthoryear{{Argiroffi} et~al.,}{{Argiroffi}
  et~al.}{2019}]{2019NatAs...3..742A}
{Argiroffi} C.,  et~al., 2019, \mn@doi [Nature Astronomy]
  {10.1038/s41550-019-0781-4}, \href
  {https://ui.adsabs.harvard.edu/abs/2019NatAs...3..742A} {3, 742}

\bibitem[\protect\citeauthoryear{{Barenfeld}, {Bubar}, {Mamajek}  \&
  {Young}}{{Barenfeld} et~al.}{2013}]{2013ApJ...766....6B}
{Barenfeld} S.~A.,  {Bubar} E.~J.,  {Mamajek} E.~E.,   {Young} P.~A.,  2013,
  \mn@doi [\apj] {10.1088/0004-637X/766/1/6}, \href
  {https://ui.adsabs.harvard.edu/abs/2013ApJ...766....6B} {766, 6}

\bibitem[\protect\citeauthoryear{{Cang} et~al.,}{{Cang}
  et~al.}{2020}]{2020A&A...643A..39C}
{Cang} T.~Q.,  et~al., 2020, \mn@doi [\aap] {10.1051/0004-6361/202037693},
  \href {https://ui.adsabs.harvard.edu/abs/2020A&A...643A..39C} {643, A39}

\bibitem[\protect\citeauthoryear{{Chen}, {Tian}, {Li}, {Wang}, {Lu}, {Xu},
  {Hou}  \& {Wu}}{{Chen} et~al.}{2022}]{2022ApJ...933...92C}
{Chen} H.,  {Tian} H.,  {Li} H.,  {Wang} J.,  {Lu} H.,  {Xu} Y.,  {Hou} Z.,
  {Wu} Y.,  2022, \mn@doi [\apj] {10.3847/1538-4357/ac739b}, \href
  {https://ui.adsabs.harvard.edu/abs/2022ApJ...933...92C} {933, 92}

\bibitem[\protect\citeauthoryear{{Cliver} \& {Dietrich}}{{Cliver} \&
  {Dietrich}}{2013}]{2013JSWSC...3A..31C}
{Cliver} E.~W.,  {Dietrich} W.~F.,  2013, \mn@doi [Journal of Space Weather and
  Space Climate] {10.1051/swsc/2013053}, \href
  {https://ui.adsabs.harvard.edu/abs/2013JSWSC...3A..31C} {3, A31}

\bibitem[\protect\citeauthoryear{{Cohen}, {Drake}, {Kashyap}, {Hussain}  \&
  {Gombosi}}{{Cohen} et~al.}{2010}]{2010ApJ...721...80C}
{Cohen} O.,  {Drake} J.~J.,  {Kashyap} V.~L.,  {Hussain} G.~A.~J.,   {Gombosi}
  T.~I.,  2010, \mn@doi [\apj] {10.1088/0004-637X/721/1/80}, \href
  {https://ui.adsabs.harvard.edu/abs/2010ApJ...721...80C} {721, 80}

\bibitem[\protect\citeauthoryear{{Collier Cameron} \& {Unruh}}{{Collier
  Cameron} \& {Unruh}}{1994}]{1994MNRAS.269..814C}
{Collier Cameron} A.,  {Unruh} Y.~C.,  1994, \mn@doi [\mnras]
  {10.1093/mnras/269.3.814}, \href
  {https://ui.adsabs.harvard.edu/abs/1994MNRAS.269..814C} {269, 814}

\bibitem[\protect\citeauthoryear{{Crosley} et~al.,}{{Crosley}
  et~al.}{2016}]{2016ApJ...830...24C}
{Crosley} M.~K.,  et~al., 2016, \mn@doi [\apj] {10.3847/0004-637X/830/1/24},
  \href {https://ui.adsabs.harvard.edu/abs/2016ApJ...830...24C} {830, 24}

\bibitem[\protect\citeauthoryear{Deng \& Welsch}{Deng \&
  Welsch}{2017}]{deng2017roles}
Deng M.,  Welsch B.~T.,  2017, Solar Physics, 292, 1

\bibitem[\protect\citeauthoryear{{Donati} \& {Landstreet}}{{Donati} \&
  {Landstreet}}{2009}]{2009ARA&A..47..333D}
{Donati} J.~F.,  {Landstreet} J.~D.,  2009, \mn@doi [\araa]
  {10.1146/annurev-astro-082708-101833}, \href
  {https://ui.adsabs.harvard.edu/abs/2009ARA&A..47..333D} {47, 333}

\bibitem[\protect\citeauthoryear{{Donati}, {Collier Cameron}, {Hussain}  \&
  {Semel}}{{Donati} et~al.}{1999}]{1999MNRAS.302..437D}
{Donati} J.~F.,  {Collier Cameron} A.,  {Hussain} G.~A.~J.,   {Semel} M.,
  1999, \mn@doi [\mnras] {10.1046/j.1365-8711.1999.02095.x}, \href
  {https://ui.adsabs.harvard.edu/abs/1999MNRAS.302..437D} {302, 437}

\bibitem[\protect\citeauthoryear{{Drake}, {Cohen}, {Yashiro}  \&
  {Gopalswamy}}{{Drake} et~al.}{2013}]{2013ApJ...764..170D}
{Drake} J.~J.,  {Cohen} O.,  {Yashiro} S.,   {Gopalswamy} N.,  2013, \mn@doi
  [\apj] {10.1088/0004-637X/764/2/170}, \href
  {https://ui.adsabs.harvard.edu/abs/2013ApJ...764..170D} {764, 170}

\bibitem[\protect\citeauthoryear{{Evensberget}, {Carter}, {Marsden},
  {Brookshaw}  \& {Folsom}}{{Evensberget}
  et~al.}{2021}]{Evensberget_Carter_Marsden_Brookshaw_Folsom_2021}
{Evensberget} D.,  {Carter} B.~D.,  {Marsden} S.~C.,  {Brookshaw} L.,
  {Folsom} C.~P.,  2021, \mn@doi [\mnras] {10.1093/mnras/stab1696}, \href
  {https://ui.adsabs.harvard.edu/abs/2021MNRAS.506.2309E} {506, 2309}

\bibitem[\protect\citeauthoryear{{Evensberget}, {Carter}, {Marsden},
  {Brookshaw}, {Folsom}  \& {Salmeron}}{{Evensberget}
  et~al.}{2022}]{evensberget2022}
{Evensberget} D.,  {Carter} B.~D.,  {Marsden} S.~C.,  {Brookshaw} L.,  {Folsom}
  C.~P.,   {Salmeron} R.,  2022, \mn@doi [\mnras] {10.1093/mnras/stab3557},
  \href {https://ui.adsabs.harvard.edu/abs/2022MNRAS.510.5226E} {510, 5226}

\bibitem[\protect\citeauthoryear{{Evensberget} et~al.,}{{Evensberget}
  et~al.}{2023}]{2023MNRAS.524.2042E}
{Evensberget} D.,  et~al., 2023, \mn@doi [\mnras] {10.1093/mnras/stad1650},
  \href {https://ui.adsabs.harvard.edu/abs/2023MNRAS.524.2042E} {524, 2042}

\bibitem[\protect\citeauthoryear{{Gopalswamy}, {Akiyama}, {Yashiro}  \&
  {M{\"a}kel{\"a}}}{{Gopalswamy} et~al.}{2010}]{2010ASSP...19..289G}
{Gopalswamy} N.,  {Akiyama} S.,  {Yashiro} S.,   {M{\"a}kel{\"a}} P.,  2010, in
  Magnetic Coupling between the Interior and Atmosphere of the Sun. pp 289--307
  (\mn@eprint {arXiv} {0903.1087}), \mn@doi{10.1007/978-3-642-02859-5_24}

\bibitem[\protect\citeauthoryear{{Gopalswamy}, {Yashiro}  \&
  {Akiyama}}{{Gopalswamy} et~al.}{2015}]{2015ApJ...809..106G}
{Gopalswamy} N.,  {Yashiro} S.,   {Akiyama} S.,  2015, \mn@doi [\apj]
  {10.1088/0004-637X/809/1/106}, \href
  {https://ui.adsabs.harvard.edu/abs/2015ApJ...809..106G} {809, 106}

\bibitem[\protect\citeauthoryear{{G{\"u}del}}{{G{\"u}del}}{2004}]{2004A&ARv..12...71G}
{G{\"u}del} M.,  2004, \mn@doi [\aapr] {10.1007/s00159-004-0023-2}, \href
  {http://adsabs.harvard.edu/abs/2004A%26ARv..12...71G} {12, 71}

\bibitem[\protect\citeauthoryear{{G{\"u}del}, {Audard}, {Reale}, {Skinner}  \&
  {Linsky}}{{G{\"u}del} et~al.}{2004}]{2004A&A...416..713G}
{G{\"u}del} M.,  {Audard} M.,  {Reale} F.,  {Skinner} S.~L.,   {Linsky} J.~L.,
  2004, \mn@doi [\aap] {10.1051/0004-6361:20031471}, \href
  {https://ui.adsabs.harvard.edu/abs/2004A&A...416..713G} {416, 713}

\bibitem[\protect\citeauthoryear{{Harra} et~al.,}{{Harra}
  et~al.}{2016}]{2016SoPh..291.1761H}
{Harra} L.~K.,  et~al., 2016, \mn@doi [\solphys] {10.1007/s11207-016-0923-0},
  \href {https://ui.adsabs.harvard.edu/abs/2016SoPh..291.1761H} {291, 1761}

\bibitem[\protect\citeauthoryear{{Hawley}, {Davenport}, {Kowalski},
  {Wisniewski}, {Hebb}, {Deitrick}  \& {Hilton}}{{Hawley}
  et~al.}{2014}]{2014ApJ...797..121H}
{Hawley} S.~L.,  {Davenport} J. R.~A.,  {Kowalski} A.~F.,  {Wisniewski} J.~P.,
  {Hebb} L.,  {Deitrick} R.,   {Hilton} E.~J.,  2014, \mn@doi [\apj]
  {10.1088/0004-637X/797/2/121}, \href
  {https://ui.adsabs.harvard.edu/abs/2014ApJ...797..121H} {797, 121}

\bibitem[\protect\citeauthoryear{{Hunter}}{{Hunter}}{2007}]{2007CSE.....9...90H}
{Hunter} J.~D.,  2007, \mn@doi [Computing in Science and Engineering]
  {10.1109/MCSE.2007.55}, \href
  {https://ui.adsabs.harvard.edu/abs/2007CSE.....9...90H} {9, 90}

\bibitem[\protect\citeauthoryear{{Hussain} et~al.,}{{Hussain}
  et~al.}{2007}]{2007MNRAS.377.1488H}
{Hussain} G.~A.~J.,  et~al., 2007, \mn@doi [\mnras]
  {10.1111/j.1365-2966.2007.11692.x}, \href
  {https://ui.adsabs.harvard.edu/abs/2007MNRAS.377.1488H} {377, 1488}

\bibitem[\protect\citeauthoryear{{Jardine} \& {Collier Cameron}}{{Jardine} \&
  {Collier Cameron}}{2019}]{2019MNRAS.482.2853J}
{Jardine} M.,  {Collier Cameron} A.,  2019, \mn@doi [\mnras]
  {10.1093/mnras/sty2872}, \href
  {https://ui.adsabs.harvard.edu/abs/2019MNRAS.482.2853J} {482, 2853}

\bibitem[\protect\citeauthoryear{{Jin} et~al.,}{{Jin}
  et~al.}{2013}]{2013ApJ...773...50J}
{Jin} M.,  et~al., 2013, \mn@doi [\apj] {10.1088/0004-637X/773/1/50}, \href
  {https://ui.adsabs.harvard.edu/abs/2013ApJ...773...50J} {773, 50}

\bibitem[\protect\citeauthoryear{{Jin}, {Manchester}, {van der Holst},
  {Sokolov}, {Toth}, {Vourlidas}, {de Koning}  \& {Gombosi}}{{Jin}
  et~al.}{2017a}]{2017ApJ...834..172J}
{Jin} M.,  {Manchester} W.~B.,  {van der Holst} B.,  {Sokolov} I.,  {Toth} G.,
  {Vourlidas} A.,  {de Koning} C.~A.,   {Gombosi} T.~I.,  2017a, \mn@doi [\apj]
  {10.3847/1538-4357/834/2/172}, \href
  {https://ui.adsabs.harvard.edu/abs/2017ApJ...834..172J} {834, 172}

\bibitem[\protect\citeauthoryear{{Jin} et~al.,}{{Jin}
  et~al.}{2017b}]{2017ApJ...834..173J}
{Jin} M.,  et~al., 2017b, \mn@doi [\apj] {10.3847/1538-4357/834/2/173}, \href
  {https://ui.adsabs.harvard.edu/abs/2017ApJ...834..173J} {834, 173}

\bibitem[\protect\citeauthoryear{{Kavanagh}, {Vidotto}, {Klein}, {Jardine},
  {Donati}  \& {{\'O} Fionnag{\'a}in}}{{Kavanagh}
  et~al.}{2021}]{2021MNRAS.504.1511K}
{Kavanagh} R.~D.,  {Vidotto} A.~A.,  {Klein} B.,  {Jardine} M.~M.,  {Donati}
  J.-F.,   {{\'O} Fionnag{\'a}in} D.,  2021, \mn@doi [\mnras]
  {10.1093/mnras/stab929}, \href
  {https://ui.adsabs.harvard.edu/abs/2021MNRAS.504.1511K} {504, 1511}

\bibitem[\protect\citeauthoryear{{Kowalski}}{{Kowalski}}{2024}]{2024LRSP...21....1K}
{Kowalski} A.~F.,  2024, \mn@doi [Living Reviews in Solar Physics]
  {10.1007/s41116-024-00039-4}, \href
  {https://ui.adsabs.harvard.edu/abs/2024LRSP...21....1K} {21, 1}

\bibitem[\protect\citeauthoryear{{Leitzinger} et~al.,}{{Leitzinger}
  et~al.}{2020}]{2020MNRAS.493.4570L}
{Leitzinger} M.,  et~al., 2020, \mn@doi [\mnras] {10.1093/mnras/staa504}, \href
  {https://ui.adsabs.harvard.edu/abs/2020MNRAS.493.4570L} {493, 4570}

\bibitem[\protect\citeauthoryear{{Lin}, {Wang}, {Deng}, {Deng}, {Mei}  \&
  {Xie}}{{Lin} et~al.}{2022}]{2022ApJ...932...62L}
{Lin} J.,  {Wang} F.,  {Deng} L.,  {Deng} H.,  {Mei} Y.,   {Xie} Y.,  2022,
  \mn@doi [\apj] {10.3847/1538-4357/ac6f54}, \href
  {https://ui.adsabs.harvard.edu/abs/2022ApJ...932...62L} {932, 62}

\bibitem[\protect\citeauthoryear{{Loesch}, {Opher}, {Alves}, {Evans}  \&
  {Manchester}}{{Loesch} et~al.}{2011}]{2011JGRA..116.4106L}
{Loesch} C.,  {Opher} M.,  {Alves} M.~V.,  {Evans} R.~M.,   {Manchester} W.~B.,
   2011, \mn@doi [Journal of Geophysical Research (Space Physics)]
  {10.1029/2010JA015582}, \href
  {https://ui.adsabs.harvard.edu/abs/2011JGRA..116.4106L} {116, A04106}

\bibitem[\protect\citeauthoryear{{Loyd} et~al.,}{{Loyd}
  et~al.}{2022}]{2022ApJ...936..170L}
{Loyd} R.~O.~P.,  et~al., 2022, \mn@doi [\apj] {10.3847/1538-4357/ac80c1},
  \href {https://ui.adsabs.harvard.edu/abs/2022ApJ...936..170L} {936, 170}

\bibitem[\protect\citeauthoryear{{Lynch}, {Airapetian}, {DeVore}, {Kazachenko},
  {L{\"u}ftinger}, {Kochukhov}, {Ros{\'e}n}  \& {Abbett}}{{Lynch}
  et~al.}{2019}]{2019ApJ...880...97L}
{Lynch} B.~J.,  {Airapetian} V.~S.,  {DeVore} C.~R.,  {Kazachenko} M.~D.,
  {L{\"u}ftinger} T.,  {Kochukhov} O.,  {Ros{\'e}n} L.,   {Abbett} W.~P.,
  2019, \mn@doi [\apj] {10.3847/1538-4357/ab287e}, \href
  {https://ui.adsabs.harvard.edu/abs/2019ApJ...880...97L} {880, 97}

\bibitem[\protect\citeauthoryear{{Lynch} et~al.,}{{Lynch}
  et~al.}{2023}]{2023BAAS...55c.254L}
{Lynch} B.~J.,  et~al., 2023, in Bulletin of the American Astronomical Society.
  p.~254 (\mn@eprint {arXiv} {2210.06476}), \mn@doi{10.3847/25c2cfeb.2dd884d5}

\bibitem[\protect\citeauthoryear{{Maehara} et~al.,}{{Maehara}
  et~al.}{2012}]{2012Natur.485..478M}
{Maehara} H.,  et~al., 2012, \mn@doi [\nat] {10.1038/nature11063}, \href
  {https://ui.adsabs.harvard.edu/abs/2012Natur.485..478M} {485, 478}

\bibitem[\protect\citeauthoryear{{Maehara}, {Shibayama}, {Notsu}, {Notsu},
  {Honda}, {Nogami}  \& {Shibata}}{{Maehara}
  et~al.}{2015}]{2015EP&S...67...59M}
{Maehara} H.,  {Shibayama} T.,  {Notsu} Y.,  {Notsu} S.,  {Honda} S.,  {Nogami}
  D.,   {Shibata} K.,  2015, \mn@doi [Earth, Planets and Space]
  {10.1186/s40623-015-0217-z}, \href
  {https://ui.adsabs.harvard.edu/abs/2015EP&S...67...59M} {67, 59}

\bibitem[\protect\citeauthoryear{{Manchester} IV et~al.,}{{Manchester}
  et~al.}{2008}]{2008ApJ...684.1448M}
{Manchester} IV W.~B.,  et~al., 2008, \mn@doi [\apj] {10.1086/590231}, \href
  {http://adsabs.harvard.edu/abs/2008ApJ...684.1448M} {684, 1448}

\bibitem[\protect\citeauthoryear{{Mason}, {Woods}, {Webb}, {Thompson},
  {Colaninno}  \& {Vourlidas}}{{Mason} et~al.}{2016}]{2016ApJ...830...20M}
{Mason} J.~P.,  {Woods} T.~N.,  {Webb} D.~F.,  {Thompson} B.~J.,  {Colaninno}
  R.~C.,   {Vourlidas} A.,  2016, \mn@doi [\apj] {10.3847/0004-637X/830/1/20},
  \href {https://ui.adsabs.harvard.edu/abs/2016ApJ...830...20M} {830, 20}

\bibitem[\protect\citeauthoryear{{Moschou}, {Drake}, {Cohen},
  {Alvarado-G{\'o}mez}, {Garraffo}  \& {Fraschetti}}{{Moschou}
  et~al.}{2019}]{2019ApJ...877..105M}
{Moschou} S.-P.,  {Drake} J.~J.,  {Cohen} O.,  {Alvarado-G{\'o}mez} J.~D.,
  {Garraffo} C.,   {Fraschetti} F.,  2019, \mn@doi [\apj]
  {10.3847/1538-4357/ab1b37}, \href
  {https://ui.adsabs.harvard.edu/abs/2019ApJ...877..105M} {877, 105}

\bibitem[\protect\citeauthoryear{{Namekata} et~al.,}{{Namekata}
  et~al.}{2017}]{2017ApJ...851...91N}
{Namekata} K.,  et~al., 2017, \mn@doi [\apj] {10.3847/1538-4357/aa9b34}, \href
  {https://ui.adsabs.harvard.edu/abs/2017ApJ...851...91N} {851, 91}

\bibitem[\protect\citeauthoryear{{Namekata} et~al.,}{{Namekata}
  et~al.}{2021}]{2022NatAs...6..241N}
{Namekata} K.,  et~al., 2021, \mn@doi [Nature Astronomy]
  {10.1038/s41550-021-01532-8}, \href
  {https://ui.adsabs.harvard.edu/abs/2022NatAs...6..241N} {6, 241}

\bibitem[\protect\citeauthoryear{{Namekata}, {Maehara}, {Honda}, {Notsu},
  {Nogami}  \& {Shibata}}{{Namekata} et~al.}{2022}]{2022arXiv221105506N}
{Namekata} K.,  {Maehara} H.,  {Honda} S.,  {Notsu} Y.,  {Nogami} D.,
  {Shibata} K.,  2022, \mn@doi [arXiv e-prints] {10.48550/arXiv.2211.05506},
  \href {https://ui.adsabs.harvard.edu/abs/2022arXiv221105506N} {p.
  arXiv:2211.05506}

\bibitem[\protect\citeauthoryear{{Namekata} et~al.,}{{Namekata}
  et~al.}{2024}]{2024ApJ...961...23N}
{Namekata} K.,  et~al., 2024, \mn@doi [\apj] {10.3847/1538-4357/ad0b7c}, \href
  {https://ui.adsabs.harvard.edu/abs/2024ApJ...961...23N} {961, 23}

\bibitem[\protect\citeauthoryear{{{\'O} Fionnag{\'a}in}, {Vidotto}, {Petit},
  {Neiner}, {Manchester}, {Folsom}  \& {Hallinan}}{{{\'O} Fionnag{\'a}in}
  et~al.}{2021}]{2021MNRAS.500.3438O}
{{\'O} Fionnag{\'a}in} D.,  {Vidotto} A.~A.,  {Petit} P.,  {Neiner} C.,
  {Manchester} W. I.,  {Folsom} C.~P.,   {Hallinan} G.,  2021, \mn@doi [\mnras]
  {10.1093/mnras/staa3468}, \href
  {https://ui.adsabs.harvard.edu/abs/2021MNRAS.500.3438O} {500, 3438}

\bibitem[\protect\citeauthoryear{{{\'O} Fionnag{\'a}in}, {Kavanagh}, {Vidotto},
  {Jeffers}, {Petit}, {Marsden}, {Morin}  \& {Golden}}{{{\'O} Fionnag{\'a}in}
  et~al.}{2022}]{2022ApJ...924..115O}
{{\'O} Fionnag{\'a}in} D.,  {Kavanagh} R.~D.,  {Vidotto} A.~A.,  {Jeffers}
  S.~V.,  {Petit} P.,  {Marsden} S.,  {Morin} J.,   {Golden} A.,  2022, \mn@doi
  [\apj] {10.3847/1538-4357/ac35de}, \href
  {https://ui.adsabs.harvard.edu/abs/2022ApJ...924..115O} {924, 115}

\bibitem[\protect\citeauthoryear{{Odert}, {Leitzinger}, {Guenther}  \&
  {Heinzel}}{{Odert} et~al.}{2020}]{2020MNRAS.494.3766O}
{Odert} P.,  {Leitzinger} M.,  {Guenther} E.~W.,   {Heinzel} P.,  2020, \mn@doi
  [\mnras] {10.1093/mnras/staa1021}, \href
  {https://ui.adsabs.harvard.edu/abs/2020MNRAS.494.3766O} {494, 3766}

\bibitem[\protect\citeauthoryear{{Osten} \& {Wolk}}{{Osten} \&
  {Wolk}}{2015}]{2015ApJ...809...79O}
{Osten} R.~A.,  {Wolk} S.~J.,  2015, \mn@doi [\apj]
  {10.1088/0004-637X/809/1/79}, \href
  {https://ui.adsabs.harvard.edu/abs/2015ApJ...809...79O} {809, 79}

\bibitem[\protect\citeauthoryear{{Powell}, {Roe}, {Linde}, {Gombosi}  \& {De
  Zeeuw}}{{Powell} et~al.}{1999}]{1999JCoPh.154..284P}
{Powell} K.~G.,  {Roe} P.~L.,  {Linde} T.~J.,  {Gombosi} T.~I.,   {De Zeeuw}
  D.~L.,  1999, \mn@doi [Journal of Computational Physics]
  {10.1006/jcph.1999.6299}, \href
  {https://ui.adsabs.harvard.edu/abs/1999JCoPh.154..284P} {154, 284}

\bibitem[\protect\citeauthoryear{{Roussev}, {Forbes}, {Gombosi}, {Sokolov},
  {DeZeeuw}  \& {Birn}}{{Roussev} et~al.}{2003}]{2003ApJ...588L..45R}
{Roussev} I.~I.,  {Forbes} T.~G.,  {Gombosi} T.~I.,  {Sokolov} I.~V.,
  {DeZeeuw} D.~L.,   {Birn} J.,  2003, \mn@doi [\apjl] {10.1086/375442}, \href
  {https://ui.adsabs.harvard.edu/abs/2003ApJ...588L..45R} {588, L45}

\bibitem[\protect\citeauthoryear{{Salas-Matamoros} \&
  {Klein}}{{Salas-Matamoros} \& {Klein}}{2015}]{2015SoPh..290.1337S}
{Salas-Matamoros} C.,  {Klein} K.~L.,  2015, \mn@doi [\solphys]
  {10.1007/s11207-015-0677-0}, \href
  {https://ui.adsabs.harvard.edu/abs/2015SoPh..290.1337S} {290, 1337}

\bibitem[\protect\citeauthoryear{{Semel}}{{Semel}}{1989}]{1989A&A...225..456S}
{Semel} M.,  1989, \aap, \href
  {https://ui.adsabs.harvard.edu/abs/1989A&A...225..456S} {225, 456}

\bibitem[\protect\citeauthoryear{{Shibata} \& {Yokoyama}}{{Shibata} \&
  {Yokoyama}}{1999}]{1999ApJ...526L..49S}
{Shibata} K.,  {Yokoyama} T.,  1999, \mn@doi [\apjl] {10.1086/312354}, \href
  {https://ui.adsabs.harvard.edu/abs/1999ApJ...526L..49S} {526, L49}

\bibitem[\protect\citeauthoryear{{Shibata} \& {Yokoyama}}{{Shibata} \&
  {Yokoyama}}{2002}]{2002ApJ...577..422S}
{Shibata} K.,  {Yokoyama} T.,  2002, \mn@doi [\apj] {10.1086/342141}, \href
  {https://ui.adsabs.harvard.edu/abs/2002ApJ...577..422S} {577, 422}

\bibitem[\protect\citeauthoryear{{Shibata} et~al.,}{{Shibata}
  et~al.}{2013}]{2013PASJ...65...49S}
{Shibata} K.,  et~al., 2013, \mn@doi [\pasj] {10.1093/pasj/65.3.49}, \href
  {https://ui.adsabs.harvard.edu/abs/2013PASJ...65...49S} {65, 49}

\bibitem[\protect\citeauthoryear{{Sokolov} et~al.,}{{Sokolov}
  et~al.}{2013}]{2013ApJ...764...23S}
{Sokolov} I.~V.,  et~al., 2013, \mn@doi [\apj] {10.1088/0004-637X/764/1/23},
  \href {https://ui.adsabs.harvard.edu/abs/2013ApJ...764...23S} {764, 23}

\bibitem[\protect\citeauthoryear{{Strassmeier}}{{Strassmeier}}{2009}]{2009A&ARv..17..251S}
{Strassmeier} K.~G.,  2009, \mn@doi [\aapr] {10.1007/s00159-009-0020-6}, \href
  {https://ui.adsabs.harvard.edu/abs/2009A&ARv..17..251S} {17, 251}

\bibitem[\protect\citeauthoryear{{Sun}, {T{\"o}r{\"o}k}  \& {DeRosa}}{{Sun}
  et~al.}{2022}]{2022MNRAS.509.5075S}
{Sun} X.,  {T{\"o}r{\"o}k} T.,   {DeRosa} M.~L.,  2022, \mn@doi [\mnras]
  {10.1093/mnras/stab3249}, \href
  {https://ui.adsabs.harvard.edu/abs/2022MNRAS.509.5075S} {509, 5075}

\bibitem[\protect\citeauthoryear{{Takahashi}, {Mizuno}  \&
  {Shibata}}{{Takahashi} et~al.}{2016}]{2016ApJ...833L...8T}
{Takahashi} T.,  {Mizuno} Y.,   {Shibata} K.,  2016, \mn@doi [\apjl]
  {10.3847/2041-8205/833/1/L8}, \href
  {https://ui.adsabs.harvard.edu/abs/2016ApJ...833L...8T} {833, L8}

\bibitem[\protect\citeauthoryear{{Titov} \& {D{\'e}moulin}}{{Titov} \&
  {D{\'e}moulin}}{1999}]{1999A&A...351..707T}
{Titov} V.~S.,  {D{\'e}moulin} P.,  1999, \aap, \href
  {https://ui.adsabs.harvard.edu/abs/1999A&A...351..707T} {351, 707}

\bibitem[\protect\citeauthoryear{{T{\'o}th} et~al.,}{{T{\'o}th}
  et~al.}{2012}]{2012JCoPh.231..870T}
{T{\'o}th} G.,  et~al., 2012, \mn@doi [Journal of Computational Physics]
  {10.1016/j.jcp.2011.02.006}, \href
  {https://ui.adsabs.harvard.edu/abs/2012JCoPh.231..870T} {231, 870}

\bibitem[\protect\citeauthoryear{Tschernitz, Veronig, Thalmann, Hinterreiter
  \& P{\"o}tzi}{Tschernitz et~al.}{2018}]{tschernitz2018reconnection}
Tschernitz J.,  Veronig A.~M.,  Thalmann J.~K.,  Hinterreiter J.,   P{\"o}tzi
  W.,  2018, The Astrophysical Journal, 853, 41

\bibitem[\protect\citeauthoryear{{Veronig}, {Odert}, {Leitzinger}, {Dissauer},
  {Fleck}  \& {Hudson}}{{Veronig} et~al.}{2021}]{2021NatAs...5..697V}
{Veronig} A.~M.,  {Odert} P.,  {Leitzinger} M.,  {Dissauer} K.,  {Fleck} N.~C.,
    {Hudson} H.~S.,  2021, \mn@doi [Nature Astronomy]
  {10.1038/s41550-021-01345-9}, \href
  {https://ui.adsabs.harvard.edu/abs/2021NatAs...5..697V} {5, 697}

\bibitem[\protect\citeauthoryear{{Vida}, {Leitzinger}, {Kriskovics}, {Seli},
  {Odert}, {Kov{\'a}cs}, {Korhonen}  \& {van Driel-Gesztelyi}}{{Vida}
  et~al.}{2019}]{2019A&A...623A..49V}
{Vida} K.,  {Leitzinger} M.,  {Kriskovics} L.,  {Seli} B.,  {Odert} P.,
  {Kov{\'a}cs} O.~E.,  {Korhonen} H.,   {van Driel-Gesztelyi} L.,  2019,
  \mn@doi [\aap] {10.1051/0004-6361/201834264}, \href
  {https://ui.adsabs.harvard.edu/abs/2019A&A...623A..49V} {623, A49}

\bibitem[\protect\citeauthoryear{{Vidotto} et~al.,}{{Vidotto}
  et~al.}{2014}]{2014MNRAS.441.2361V}
{Vidotto} A.~A.,  et~al., 2014, \mn@doi [\mnras] {10.1093/mnras/stu728}, \href
  {https://ui.adsabs.harvard.edu/abs/2014MNRAS.441.2361V} {441, 2361}

\bibitem[\protect\citeauthoryear{{Vidotto}, {Bourrier}, {Fares}, {Bellotti},
  {Donati}, {Petit}, {Hussain}  \& {Morin}}{{Vidotto}
  et~al.}{2023}]{2023A&A...678A.152V}
{Vidotto} A.~A.,  {Bourrier} V.,  {Fares} R.,  {Bellotti} S.,  {Donati} J.~F.,
  {Petit} P.,  {Hussain} G.~A.~J.,   {Morin} J.,  2023, \mn@doi [\aap]
  {10.1051/0004-6361/202347237}, \href
  {https://ui.adsabs.harvard.edu/abs/2023A&A...678A.152V} {678, A152}

\bibitem[\protect\citeauthoryear{{Virtanen} et~al.,}{{Virtanen}
  et~al.}{2020}]{2020SciPy-NMeth}
{Virtanen} P.,  et~al., 2020, \mn@doi [Nature Methods]
  {https://doi.org/10.1038/s41592-019-0686-2}, \href {https://rdcu.be/b08Wh}
  {17, 261}

\bibitem[\protect\citeauthoryear{{Waite}, {Marsden}, {Carter}, {Petit},
  {Donati}, {Jeffers}  \& {Boro Saikia}}{{Waite}
  et~al.}{2015}]{2015MNRAS.449....8W}
{Waite} I.~A.,  {Marsden} S.~C.,  {Carter} B.~D.,  {Petit} P.,  {Donati} J.~F.,
   {Jeffers} S.~V.,   {Boro Saikia} S.,  2015, \mn@doi [\mnras]
  {10.1093/mnras/stv006}, \href
  {https://ui.adsabs.harvard.edu/abs/2015MNRAS.449....8W} {449, 8}

\bibitem[\protect\citeauthoryear{Webb \& Howard}{Webb \&
  Howard}{2012}]{webb2012coronal}
Webb D.~F.,  Howard T.~A.,  2012, Living Reviews in Solar Physics, 9, 1

\bibitem[\protect\citeauthoryear{{Yashiro}, {Gopalswamy}, {Akiyama}, {Michalek}
   \& {Howard}}{{Yashiro} et~al.}{2005}]{2005JGRA..11012S05Y}
{Yashiro} S.,  {Gopalswamy} N.,  {Akiyama} S.,  {Michalek} G.,   {Howard}
  R.~A.,  2005, \mn@doi [Journal of Geophysical Research (Space Physics)]
  {10.1029/2005JA011151}, \href
  {https://ui.adsabs.harvard.edu/abs/2005JGRA..11012S05Y} {110, A12S05}

\bibitem[\protect\citeauthoryear{{Yashiro}, {Akiyama}, {Gopalswamy}  \&
  {Howard}}{{Yashiro} et~al.}{2006}]{2006ApJ...650L.143Y}
{Yashiro} S.,  {Akiyama} S.,  {Gopalswamy} N.,   {Howard} R.~A.,  2006, \mn@doi
  [\apjl] {10.1086/508876}, \href
  {https://ui.adsabs.harvard.edu/abs/2006ApJ...650L.143Y} {650, L143}

\bibitem[\protect\citeauthoryear{{van der Holst}, {Sokolov}, {Meng}, {Jin},
  {Manchester}, {T{\'o}th}  \& {Gombosi}}{{van der Holst}
  et~al.}{2014}]{2014ApJ...782...81V}
{van der Holst} B.,  {Sokolov} I.~V.,  {Meng} X.,  {Jin} M.,  {Manchester}
  W.~B. I.,  {T{\'o}th} G.,   {Gombosi} T.~I.,  2014, \mn@doi [\apj]
  {10.1088/0004-637X/782/2/81}, \href
  {https://ui.adsabs.harvard.edu/abs/2014ApJ...782...81V} {782, 81}

\bibitem[\protect\citeauthoryear{{van der Walt}, {Colbert}  \&
  {Varoquaux}}{{van der Walt} et~al.}{2011}]{2011CSE....13b..22V}
{van der Walt} S.,  {Colbert} S.~C.,   {Varoquaux} G.,  2011, \mn@doi
  [Computing in Science and Engineering] {10.1109/MCSE.2011.37}, \href
  {https://ui.adsabs.harvard.edu/abs/2011CSE....13b..22V} {13, 22}

\makeatother
\end{thebibliography}

\bsp	%
\label{lastpage}
\end{document}